\documentclass{aa}
\usepackage{graphicx}
\usepackage{txfonts}
\usepackage{natbib}
\usepackage{comment}
\usepackage{longtable}
\usepackage{array}

\bibpunct{(}{)}{;}{a}{}{,}

\begin{document}

\title{Magnetic activity in the HARPS M dwarf sample
\thanks{Table 6 is only available in electronic form at the CDS via anonymous ftp to cdsarc.u-strasbg.fr (130.79.128.5) or via http://cdsweb.u-strasbg.fr/cgi-bin/qcat?J/A+A/}
}

\subtitle{The rotation-activity relationship for very low-mass stars 
   through $R^\prime_{HK}$  
 }

\authorrunning{Astudillo-Defru et al.}
\author{N. Astudillo-Defru \inst{1, 2, 3}, X. Delfosse\inst{1, 2}, X. Bonfils\inst{1, 2}, T. Forveille\inst{1, 2}, C. Lovis\inst{3}, J. Rameau\inst{1, 2, 4}}

\institute{Univ. Grenoble Alpes, IPAG, F-38000 Grenoble, France 
  \and CNRS, IPAG, F-38000 Grenoble, France 
  \and Observatoire de Gen\`eve, Universit\'e de Gen\`eve, 51 ch. des Maillettes, 1290 Sauverny, Switzerland 
  \and D\'epartement de Physique, Universit\'e de Montr\'eal, C.P. 6128 Succ. Centre-Ville, Montr\'eal, QC H3C 3J7, Canada 
}

\date{}

\abstract
{Atmospheric magnetic fields in stars with convective envelopes heat 
stellar chromospheres, and thus increase the observed flux in the 
\ion{Ca}{ii} H and K doublet. Starting with the historical Mount Wilson 
monitoring program, these two spectral lines have been widely used to 
trace stellar magnetic activity, and as a proxy for rotation 
period ($P_{\rm rot}$) and consequently for stellar age. Monitoring 
stellar activity has also become essential in filtering out 
false-positives due to magnetic activity in extra-solar planet 
surveys. The \ion{Ca}{ii} emission is traditionally quantified 
through the $R^\prime_{HK}$-index, which compares the chromospheric 
flux in the doublet to the overall bolometric flux of the star. 
Much work has been done to characterize this index for FGK-dwarfs, 
but M dwarfs -- the most numerous stars of the Galaxy -- were left
out of these analyses and no calibration of their 
\ion{Ca}{ii} H and K emission to an $R^\prime_{HK}$ exists
to date.}
{We set out to characterize the magnetic activity of the low- and very-low-mass 
stars by providing a calibration of the $R^\prime_{HK}$-index 
that extends to the realm of M dwarfs, and by evaluating the relationship between 
$R^\prime_{HK}$ and the rotation period.}
{We calibrated 
the bolometric and photospheric factors for M dwarfs to properly 
transform the S-index (which compares the flux in the \ion{Ca}{ii} H and K lines 
to a close spectral continuum) into the $R^\prime_{HK}$.
We monitored magnetic activity through the \ion{Ca}{ii} 
H\textrm{\small \ and }K emission lines in the HARPS M dwarf sample.}
{The $R^\prime_{HK}$ index, like the fractional X-ray luminosity $L_X/L_{bol}$, 
shows a saturated correlation with rotation, with saturation setting
in around a ten days rotation period. Above that period, slower
rotators show weaker \ion{Ca}{ii} activity, as expected. Under
that period, the $R^\prime_{HK}$ index saturates to approximately $10^{-4}$. 
Stellar mass modulates the \ion{Ca}{ii} activity, 
with $R^\prime_{HK}$ showing a constant basal activity above 0.6M$_\odot$ 
and then decreasing with mass between 0.6M$_\odot$  and the 
fully-convective limit of 0.35M$_\odot$. Short-term variability 
of the activity correlates with its mean level and stars with higher 
$R^\prime_{HK}$ indexes show larger  $R^\prime_{HK}$ variability, as 
previously observed for earlier spectral types.
}
{}


\maketitle

\section{Introduction}
\label{sec:intro}

`Stellar activity' generically describes the various observational 
consequences of enhanced magnetic fields, whether those appear on the 
stellar photosphere, in the chromosphere, or in the corona. 
In low-mass stars, magnetic fields are in turn believed to originate 
from dynamo processes \citep[e.g.,][]{1955ApJ...122..293P}. Stellar 
activity is thus used as a diagnostic of the dynamo over a wide range 
of stellar ages, masses, and rotational periods. 

It is well established that a larger fraction of M dwarfs exhibit evidence of magnetic
activity than their more massive Sun-like siblings. The fraction of stars
showing $H_{\alpha}$ chromospheric emission \citep{1998A&A...331..581D} 
or frequent flare \citep{1991ApJ...378..725H, 2014ApJ...781L..24S, 2014ApJ...797..121H}
increase when the mass decreases, for two reasons.
On the one hand, 
lower-mass stars have much longer rotational braking times 
\citep{1998A&A...331..581D, 2003ApJ...586..464B, 2011MNRAS.413.2218D} 
and on the other, lower-mass stars show stronger
chromospheric and coronae emission for a given rotation period 
\citep[e.g.,][]{2007AcA....57..149K}. Spectropolarimetric 
observations demonstrate that the more massive (M$_\star > 0.5~$M$_\odot$) 
M dwarfs with reconstructed magnetic topologies have magnetic 
fields with a strong toroidal component, reminiscent of those of 
active K and G dwarfs, whereas the lowest-mass M dwarfs 
exhibit magnetic fields that are mainly poloidal 
\citep{2008MNRAS.390..567M, 2008MNRAS.390..545D}. This transition 
takes place slightly above the theoretical full-convection threshold 
(M$\sim$0.35M$_\odot$), suggesting that the dynamo mechanism might change 
when the tachocline can no longer play a major role as
the radiative core becomes negligibly small \citep{2000ARA&A..38..337C}. 
Recent theoretical works successfully reproduce such large-scale 
and small-scale field properties of magnetic fields in fully 
convective stars \citep[e.g.][]{2015ApJ...813L..31Y}.

Large-scale radial velocity searches for extra-solar planets have 
helped rejuvenate studies of stellar activity, both because 
they provide extensive time series of high-resolution spectra 
for large samples of stars and because characterizing 
the activity of a star is essential to avoid confusing symptoms 
of that activity with a planetary signal. Magnetic inhibition 
of surface convection, spots, plages, and other inhomogeneities of 
the stellar surface, indeed all affect the shape of spectral lines, 
shifting their centroids, and consequently biasing the measured radial 
velocity. This unwanted signal is often handled as random noise, part of
the so-called RV-{\it{jitter}}, and added in quadrature to the known
noise sources such as photon noise and instrumental instabilities. 
It can, however, be coherent over the $\sim$months-long time-scale 
of the stellar rotation, or over the $\sim$years-long time-scale of a 
stellar activity cycle, and can thus be mistaken for the signature 
of a planetary companion 
\citep[e.g.,][]{2001A&A...379..279Q,2007A&A...474..293B,
2014Sci...345..440R}.  This has stimulated extensive work to model 
the effect, identify proxies for its source, and filter it out of 
the RV time series
\citep{2011A&A...525A.140D,
2011A&A...527A..82D,2011A&A...528A...4B,2012A&A...545A.109B,
2010A&A...512A..38L,2010A&A...512A..39M,2013A&A...551A.101M,
2014arXiv1405.2016T}.

\begin{figure*}[t]
\centering
\resizebox{\hsize}{!}{\includegraphics{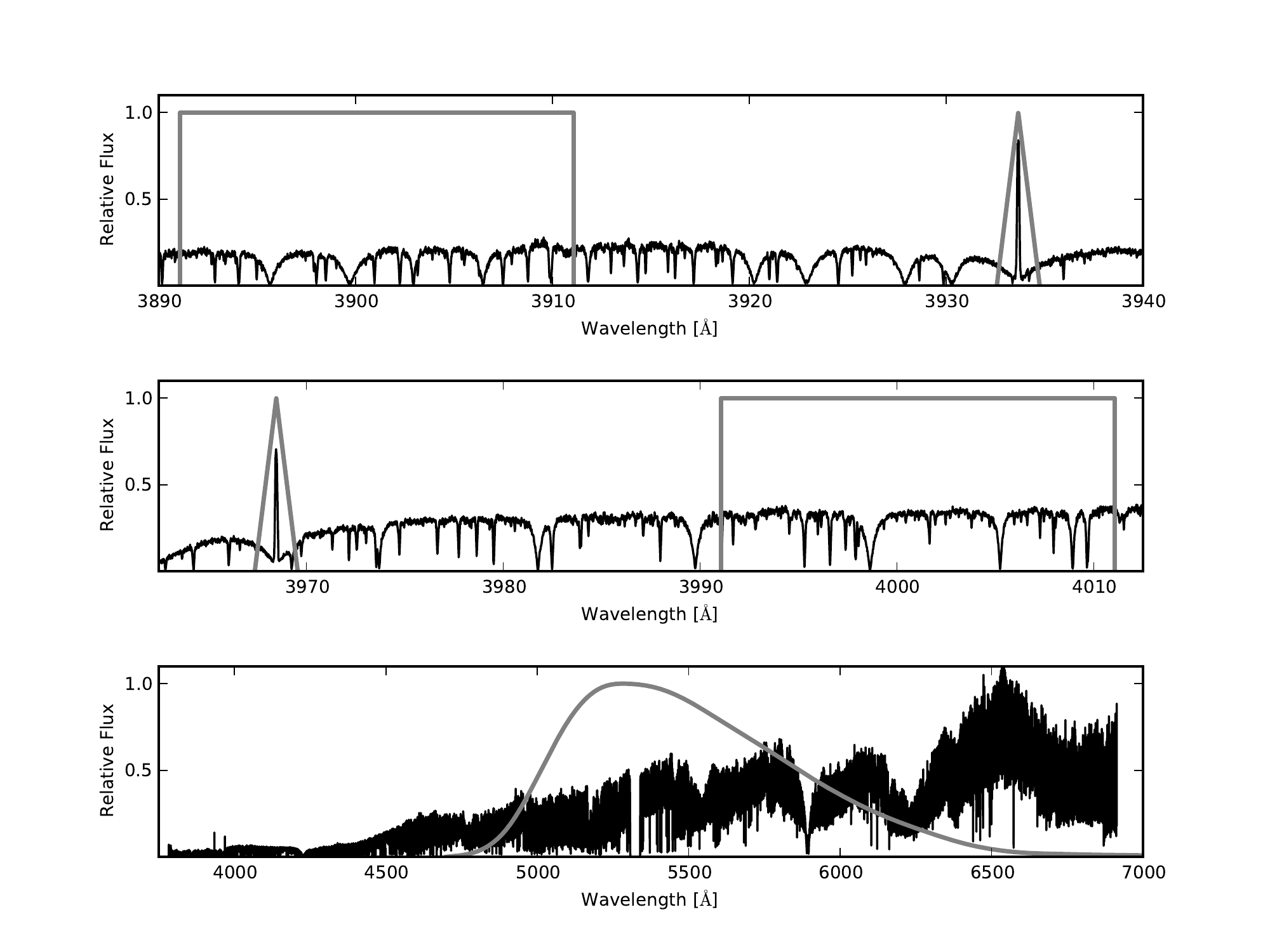}}
\caption{Median HARPS rest-frame spectrum of a representative mid-M dwarf 
(Gl~699) with overlays of the V, K (top panel), and H, R (middle panel) 
filters of the Mount Wilson system and of the photometric $V$ band 
(bottom panel) 
}
\label{fig:bands}
\end{figure*}

Emission in the core of the \ion{Ca}{ii} H and K resonance 
lines (396.8 nm and 393.4 nm) reflects non-thermal heating in the 
chromosphere that produces bright plages, and is perhaps the most widely 
used of these activity diagnostics. The historical Mount Wilson program 
\citep{1978PASP...90..267V} intensively monitored this activity proxy 
for approximately sixty solar-type stars and quantified stellar activity through 
the so-called S-index. That index is the ratio between the flux through two 
triangular band-passes (with 1.09~$\AA$ full width at half maximum (FWHM°)) centered 
on the \ion{Ca}{ii} H and K lines and the flux through two 20~$\AA$-wide 
rectangular pseudo-continuum band-passes on the violet (V, centered 
at 3901 $\AA$) and red (R, centered at 4001~$\AA$) sides of the lines 
(Fig.~\ref{fig:bands}). 

The S-index is akin to an equivalent width, and well matched to its initial
purpose of quantifying variations in the activity of a given star. It can
also be used to compare activity levels within a narrow spectral type bin, 
but is poorly suited to comparing stars of different spectral types. 
To account for the variation of the continuum level with spectral type,
\citet{1982A&A...107...31M} and \citet{1984A&A...130..353R} introduced the 
$C_{cf}$ factor, which is the ratio between the fraction of the stellar 
luminosity emitted in the \ion{Ca}{ii} H and K lines and the S-index 
(alternatively, $C_{cf}$ can be thought of as a bolometric 
correction for a photometric filter defined by the two pseudo-continuum bands
of the Mount Wilson system). $C_{cf}$ can be estimated from a broad-band 
color index, and then used to convert the S-index into the fractional 
luminosity on the \ion{Ca}{ii} filters. Calibrations of the $C_{cf}$ factor 
against a broad-band color have, to date, focused on FGK-dwarfs, with 
poor coverage of the M dwarfs. Furthermore, these calibrations tend to 
use $B-V$ as their color index, which happens to be a poor choice 
for M dwarfs: these stars emit little flux in the B band, and their 
V-flux is also sensitive to metallicity 
\citep[e.g.,][]{2000A&A...364..217D, 2005A&A...442..635B}. 
As a consequence, $C_{cf}$ plotted against $B-V$ has a large 
scatter for $B-V>1.2$ (late-K) \citep[][Fig.~3]{1980PASP...92..385V}.

The two triangular band-passes of the Mount Wilson system centered on 
the \ion{Ca}{ii} H and K lines measure a combination of
chromospheric and photospheric emissions. $R_{HK}$ is consequently a fractional
luminosity in the \ion{Ca}{ii} lines, rather than a fractional 
chromospheric luminosity in those lines.
Corrections for the photospheric contribution were first proposed by
\citet{1974A&A....33..257B} and \cite{1978ApJ...220..619L}, but 
\citet{1984ApJ...276..254H} has become the standard reference. 
They compute an S-index from the spectrum of the photosphere, 
S$_{phot}$, which the \citet{1982A&A...107...31M} color-dependent 
calibration transforms into a $R_{HK}$ for the photosphere, 
$R_{phot}$. $R^\prime_{HK}$ then results as $R^\prime_{HK} = R_{HK}-R_{phot}$. 
The activity index $R^\prime_{HK}$ is thus the fraction of the stellar bolometric 
luminosity which the chromosphere emits in the \ion{Ca}{ii} H and K lines.
Full details on the conversion of the S-index into $R^\prime_{HK}$ are 
given in the Appendix of \citet{1984ApJ...279..763N}.

The $R^\prime_{HK}$ index is well characterized for the FGK-dwarfs
\citep{1996AJ....111..439H,2000A&A...361..265S,2004ApJS..152..261W,
2007AJ....133..862H,2010ApJ...725..875I,2011arXiv1107.5325L},
and some authors \citep{2000A&AS..142..275S,2002MNRAS.332..759T,
2006MNRAS.372..163J} extrapolated the corresponding $C_{cf}$ and 
$R_{phot}$ conversion factors to later dwarfs. The
validity of such extrapolations to redder colors is 
questionable, however. Two further, recent studies proposed better grounded
calibrations of the \ion{Ca}{ii} H and K emission of M dwarfs. 
\citet{2010AJ....139..504B} determined $L_{Ca}/L_{bol}$ from
the \ion{Ca}{ii} H and K equivalent widths of a sample of M dwarfs, 
but they chose not to anchor their index on the Mount Wilson 
$R^\prime_{HK}$, complicating comparisons with solar-type stars. 
\citet{2013A&A...549A.117M} used synthetic spectra computed to provide 
conversions from $S$ to $R^\prime_{HK}$, but their use of $B-V$ as the 
color index is, as discussed above, less than ideal for M dwarfs. 
Recently, \citet{2015MNRAS.452.2745S} analyzed the stellar 
rotation against activity for F-type to mid-M dwarfs; there they 
provide extended relationships for the $C_{cf}$ and $R_{phot}$ factors 
obtained following the classical works of \citet{1982A&A...107...31M} , 
\citet{1984A&A...130..353R}, and \citet{1984ApJ...276..254H}, respectively.

In the present work, we calibrate the $C_{cf}$ and $R_{phot}$ factors
as a function of $B-V$, $V-I$, and $I-K$ for the early to mid-M dwarfs. 
Sections~\ref{sec:scaling_S}, \ref{sec:bol_cor} and \ref{sec:phot_cor} 
describe our alternative methodology to calibrate such factors. 
In section~\ref{sec:Rhk_rotation}, we then use the resulting $R^\prime_{HK}$ 
for the HARPS M dwarf sample to examine how magnetic activity depends 
on stellar rotation, while section~\ref{sec:Rhk_harps_sample} analyses 
how $R^\prime_{HK}$ varies with the overall stellar parameters.

\section{Observations and data reduction}
\label{sec:obs}

\subsection{HARPS spectra}

We used spectra observed with the high accuracy radial velocity planets 
searcher (HARPS) installed on the ESO 3.6m telescope at La Silla observatory 
in Chile. HARPS is a fiber-fed cross-dispersed echelle spectrograph covering 
the 380-690 nm spectral range with a resolving power of 115,000. The stellar 
light is injected into a science fiber and a second fiber can be illuminated 
either by a ThAr lamp for simultaneous calibration or by the sky for subtraction
of its emission \citep{2003Msngr.114...20M}.

In order to obtain a measurement of the fraction of luminosity 
that M dwarfs emit from the \ion{Ca}{ii} H and K line we make 
a spectrophotometric analysis in using the HARPS spectra.

Since our spectra of M dwarfs are not flux-calibrated, we need
to reference them to those of better characterized stars
to calibrate $C_{cf}$ and $R_{phot}$, and we use spectra of
GK-dwarfs for that role. Both the GK and the M stars were 
originally observed to search for exoplanets through 
high-precision RV monitoring \citep{2011arXiv1107.5325L, 
2012A&A...546A..27B, 2013A&A...549A.109B}, with selection criteria 
described in detail in the above papers. Briefly, the GK-dwarfs 
are within 50~pc and have low projected rotational velocity 
$v\;sin\,i<3-4\;kms^{-1}$; their spectra usually have a signal-to-noise 
(S/N) ratio above 100 per pixel at 550 nm. 
Our M dwarfs sample includes $\sim$300 M dwarfs closer than 20 pc, 
brighter than V = 12 mag and southward of $\delta$ = 15$^\circ$, 
as well as $\sim$40 fainter stars kept from the initial GTO sample 
(V $<$ 14 mag; d $<$ 11 pc; $\delta <$ 15$^\circ$; $v\;sin\,i\leq6.5\;kms^{-1}$).

The HARPS pipeline \citep{2007A&A...468.1115L} automatically 
reduces the spectroscopic images to spectra, making use of 
calibrations obtained during day time. It then extracts the 
radial velocity from the cross-correlation (CCF) of the spectrum 
and a binary mask, which also provides a FWHM, 
a contrast, and a bisector-span. In this work, we start from order-merged 
and background-subtracted spectra corrected from the motion of the 
observatory relative to the barycenter of the solar system. We
used the stellar radial velocity provided by the pipeline to
re-center those spectra to the stellar rest-frame. Finally, we 
corrected from the instrumental transmittance (normalized to unity) 
using the ratio between a BT-settle \citep{2014IAUS..299..271A} 
theoretical spectrum for the bright (V=6.88) G-dwarf HD~223171 
and the average of 20 observed spectra of that star. This 
correction therefore assumes that the transmittance does not 
vary and is consequently approximate. This, however, is of 
little consequence since, as described below, our $C_{cf}$ and 
$R_{phot}$ of M dwarfs are anchored on those of solar-type stars 
having these two parameters well-calibrated and
observed under very similar conditions.

\subsection{Literature photometry, parallaxes, and physical parameters}

We obtained BVIK photometry of the M dwarfs from \citet{1992ApJS...82..351L}, 
\citet{2014MNRAS.443.2561G} and \citet{2003yCat.2246....0C}; 
and of the GK-dwarf calibrators from \citet{2001ApJ...558..309D} 
and \citet{2003yCat.2246....0C}. When needed, we used the transformations of
\citet{2001AJ....121.2851C} to homogenize this photometry to the 
Johnson-Cousins-CIT system. 
We adopted parallaxes ($\pi$) from \citet{2007A&A...474..653V}, 
\citet{1995gcts.book.....V}, \citet{1997ESASP1200.....P}, 
\citet{1997AJ....113.1458H} and the research consortium on nearby 
stars (RECONS) parallax program \citep[e.g.,][]{2010AJ....140..897R, 2011AJ....141..117J}. 

For the GK-dwarfs, we adopt the effective temperature, radii, masses, and 
metallicities listed in \citet{2008A&A...487..373S}. For the M dwarfs  
we obtained the metallicities from \citet{2013A&A...551A..36N}, computed 
effective temperatures and radii using the \citet{2012ApJ...757..112B} 
V-K/metallicity relations, and the stellar masses using the
\citet{2000A&A...364..217D} mass vs K-band absolute magnitude relation.
The later relation is valid between 0.09M$_\odot$ and 0.7M$_\odot$, and
the few masses between 0.7M$_\odot$ and 0.8M$_\odot$ are therefore based on
a slight extrapolation.

\section{Scaling the S-index from HARPS observations}
\label{sec:scaling_S}

\begin{figure}[t]
\centering
\resizebox{\hsize}{!}{\includegraphics{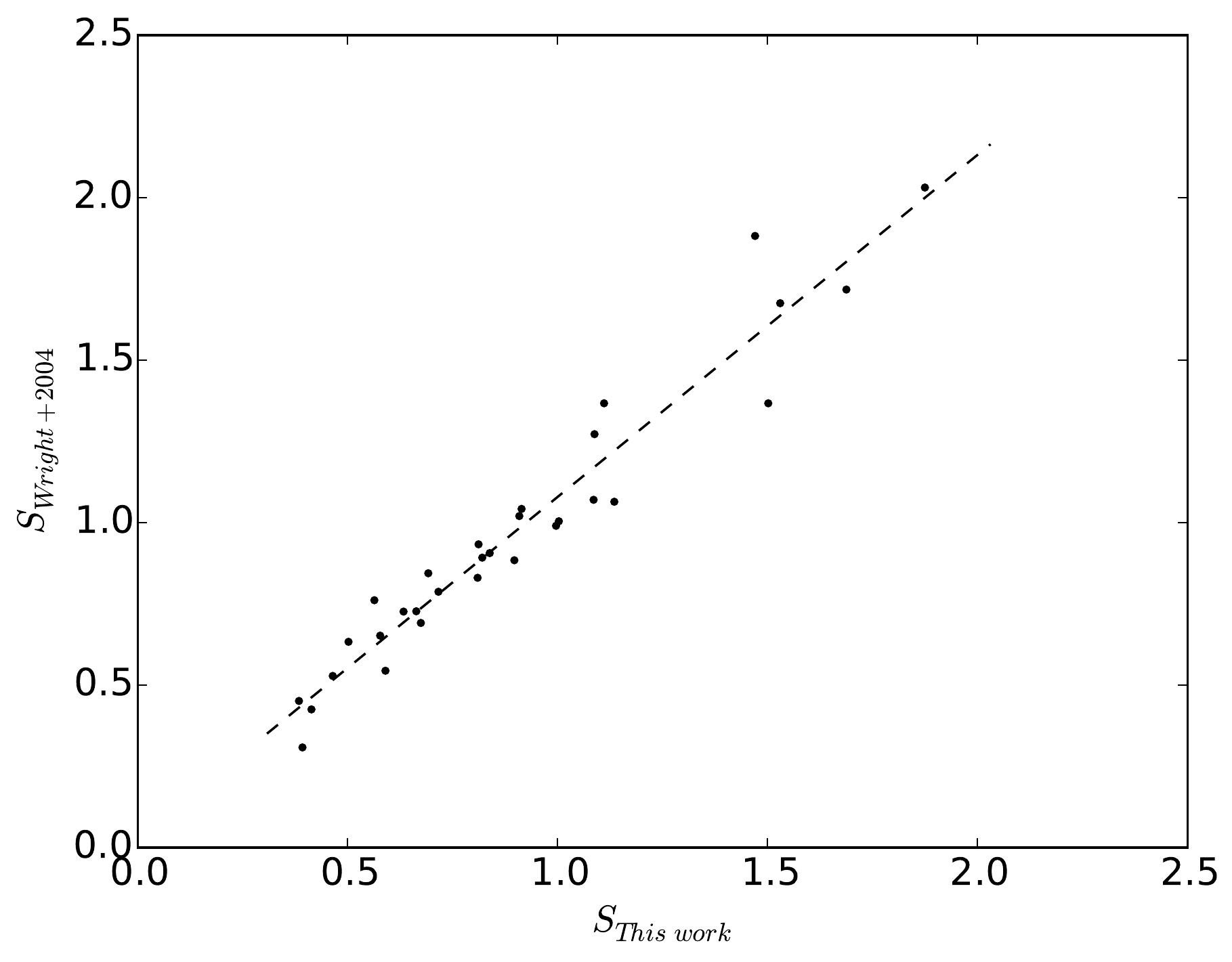}}
\caption{Median HARPS S-index against the \citet{2004ApJS..152..261W} 
S-index for the 31 targets in common. The solid line represents the 
linear least-square fit (Eq.~\ref{eq:s_harps_wright_lin}) between  the
two data sets, which we use to bring our S measurements onto the 
Mount Wilson scale.}
\label{fig:s_harps_wright}
\end{figure}

The definition of the S-index traces back to the Mount~Wilson stellar 
activity program, and modern measurements are traditionally brought
onto the scale defined by that program to ease inter-comparisons.
The long-term Mount Wilson program started with a Coud\'e scanner 
of the 100-inch telescope (HKP-1) 
\citep{1968ApJ...153..221W,1978ApJ...226..379W} 
and later transitioned to a photometer on the 60-inch telescope 
(HKP-2) \citep{1978PASP...90..267V}. 

S is defined as

\begin{equation}
\label{eq:S_definition}
S = \alpha \cdot {{f_H + f_K} \over {f_V + f_R}}
,\end{equation}

where $f_H,\;f_K,\;$ $f_V,$ and $f_R$ are the total counts in  
the four pass-bands described above (Sec.~\ref{sec:intro} and 
Fig.~\ref{fig:bands}) and $\alpha$=2.4 is a calibration constant 
that brings the 100-inch and 60-inch activity indices into 
approximate agreement.

To compute an S-index with HARPS
we chose to follow \citet{2011arXiv1107.5325L} : instead of working
with integrated flux in each passband, we used the mean flux per wavelength 
interval $\tilde{f_H} = f_H/\Delta\lambda_H$, 
$\tilde{f_K} = f_K/\Delta\lambda_K$, $\tilde{f_V} = f_V/\Delta\lambda_V$ and
$\tilde{f_R} = f_R/\Delta\lambda_R$. To be coherent with the Mount Wilson one, 
our $S$ relationship must be normalized by the ratio of the effective bandpass width 
$\Delta\lambda_H = \Delta\lambda_K = 1.09\,\AA$ 
and $\Delta\lambda_V = \Delta\lambda_R = 20\,\AA$. In addition,
we must account for the Mount 
Wilson program's exposing eight times as long in its narrow H and K bands 
than in its broader V and R bands, versus our using. The HARPS calibration constant $\alpha_H$ is therefore
\begin{displaymath}
\alpha_H =  \alpha \cdot 8 \cdot {1.09\,\AA \over 20\,\AA} \sim 1,
\end{displaymath}
and the S-index can be written as :

\begin{equation}
\label{eq:s_harps}
S \approx {{\tilde{f_H} + \tilde{f_K}} \over {\tilde{f_V} + \tilde{f_R}}}
.\end{equation}

We have no target in common with the Mount Wilson program to directly verify 
the consistency of our S values with its scale, but \citet{2004ApJS..152..261W} 
scaled their Keck and Lick S-indices to Mount Wilson measurements. We 
have 31 targets in common with them
\footnote{The common targets are Gl~465, Gl~357, Gl~1, Gl~581, Gl~87, Gl~667C, 
Gl~486, Gl~686, Gl~436, Gl~105B, Gl~699, Gl~526, Gl~433, Gl~273, 
Gl~555, Gl~628, Gl~413.1, GJ~2066, Gl~701, Gl~393, Gl~876, Gl~849, 
Gl~536, Gl~887, Gl~514, Gl~176, Gl~678.1A, Gl~229, Gl~846, Gl~880, 
and Gl~382} (Fig.~\ref{fig:s_harps_wright}), which we use to assess the 
consistency of the HARPS S-indices with the Mount Wilson scale. The best 
linear fit between between the two datasets is

\begin{equation}
\label{eq:s_harps_wright_lin}
S_{M.\, W.} = 1.053 \cdot S_{HARPS}+0.026
.\end{equation}
The uncertainties in the slope and the intercept from the covariance matrix 
are 0.0025 and 0.0024, respectively, 
while the root-mean square deviation of the residuals from that fit, 0.080, is
consistent with that expected from variations of the stellar activity 
between the two non-contemporaneous measurements 
as illustrated by the dispersion on the S-index obtained
by \citet{2004ApJS..152..261W} or those listed in our table \ref{tab:caiihkharps}.
Furthermore, the residuals from our fits are higher for large values of the S-index, which is an
expected behavior since more active stars show the largest intrinsic variability
of the S-index.
The small 1.053 factor most likely accounts for minor mismatches between
our synthetic filters and the original physical Mount Wilson bandpasses.
We use Eq.~\ref{eq:s_harps_wright_lin} to bring our S measurements 
onto the Mount Wilson scale.

$R^\prime_{HK}$ derives from an S-index on the Mount Wilson scale through
\begin{eqnarray}
\label{eq:Rhk}
R^\prime_{HK} & = & R_{HK} - R_{phot} \nonumber \\
           & = & K \cdot \sigma^{-1} \cdot 10^{-14} \cdot C_{cf} \cdot (S-S_{phot})
,\end{eqnarray}
where $R_{phot}$ and $S_{phot}$ stand for the photospheric contribution to $R$
and $S$, $C_{cf}$ is the bolometric factor described in Sec.~1, $\sigma$ is 
the Stefan-Boltzmann constant, and $10^{-14}$ is a scaling factor. $K$ converts 
the surface fluxes from arbitrary units to physical fluxes on the stellar 
surface 
\citep[for a detailed description see, e.g.,][section 2.d]{1984A&A...130..353R}. 
\citet{1982A&A...107...31M}, \citet{1984A&A...130..353R}, and 
\citet{2007AJ....133..862H} respectively find 
$K=0.76\times10^6,\, 1.29\times10^6,\, 1.07\times10^6\, [erg\,cm^{-2}\,s^{-1}]$. 
We use the later value, which is referenced to more recent solar data, hence, 
$K \cdot \sigma^{-1} \cdot 10^{-14}$=1.887$\times$10$^{-4}$.

As discussed in Sec.~1, $C_{cf}$ and $R_{phot}$ were previously poorly 
constrained in the M dwarfs domain.

\section{The bolometric factor $C_{cf}$}
\label{sec:bol_cor}

The bolometric factor is :
\begin{eqnarray}
\label{eq:ccf}
C_{cf} & \equiv & {({f_V+f_R}) \over f_{bol}} 
,\end{eqnarray}
where $f_V$ and $f_R$ are defined in Eq.~(\ref{eq:S_definition}), 
and $f_{bol}$ is the apparent bolometric flux of the star. Previous works 
directly applied Equation~(\ref{eq:ccf}) to FGK-dwarfs to derive a $C_{cf}$ -color-dependent relation \citep{1982A&A...107...31M,1984A&A...130..353R}.

Our HARPS M dwarf spectra are not flux calibrated, and chromatic variations in 
both seeing (hence fiber injection efficiency) and atmospheric transmission 
therefore prevent the absolute spectro-photometric computation of $f_V$, $f_R$ and $f_{bol}$. 
However, $C_{cf}$ being a flux ratio, its empirical determination is only important for 
the variation of the transmission (atmospheric and instrumental) in function of the wavelength,
which can be corrected using a differential method discussed below.

Since our spectra were often observed close in time, and in similar condition to G-K 
dwarfs, they can be used to bootstrap the computation through the ratio of 
the $C_{cf}$s  for the M dwarf and its G or K standard:
\begin{eqnarray}
\label{eq:ccf2}
{C_{cf,M} \over C_{cf,Std}} & = & \frac{({f_V+f_R})_M}{({f_V+f_R})_{Std}}  \frac{f_{bol,Std}}{f_{bol,M}} 
,\end{eqnarray}
where the chromatic components of both atmospheric absorption and injection 
efficiency cancel out as long as the M dwarf and its standard star were 
observed under even moderately similar atmospheric conditions. We stress 
that the use of $\tilde{f_V},\ \tilde{f_R}$ or $f_V,\ f_R$ (defined above) in 
Eq.~(\ref{eq:ccf2}) is equivalent.

\begin{figure}[tp]
\centering
\resizebox{\hsize}{!}{\includegraphics{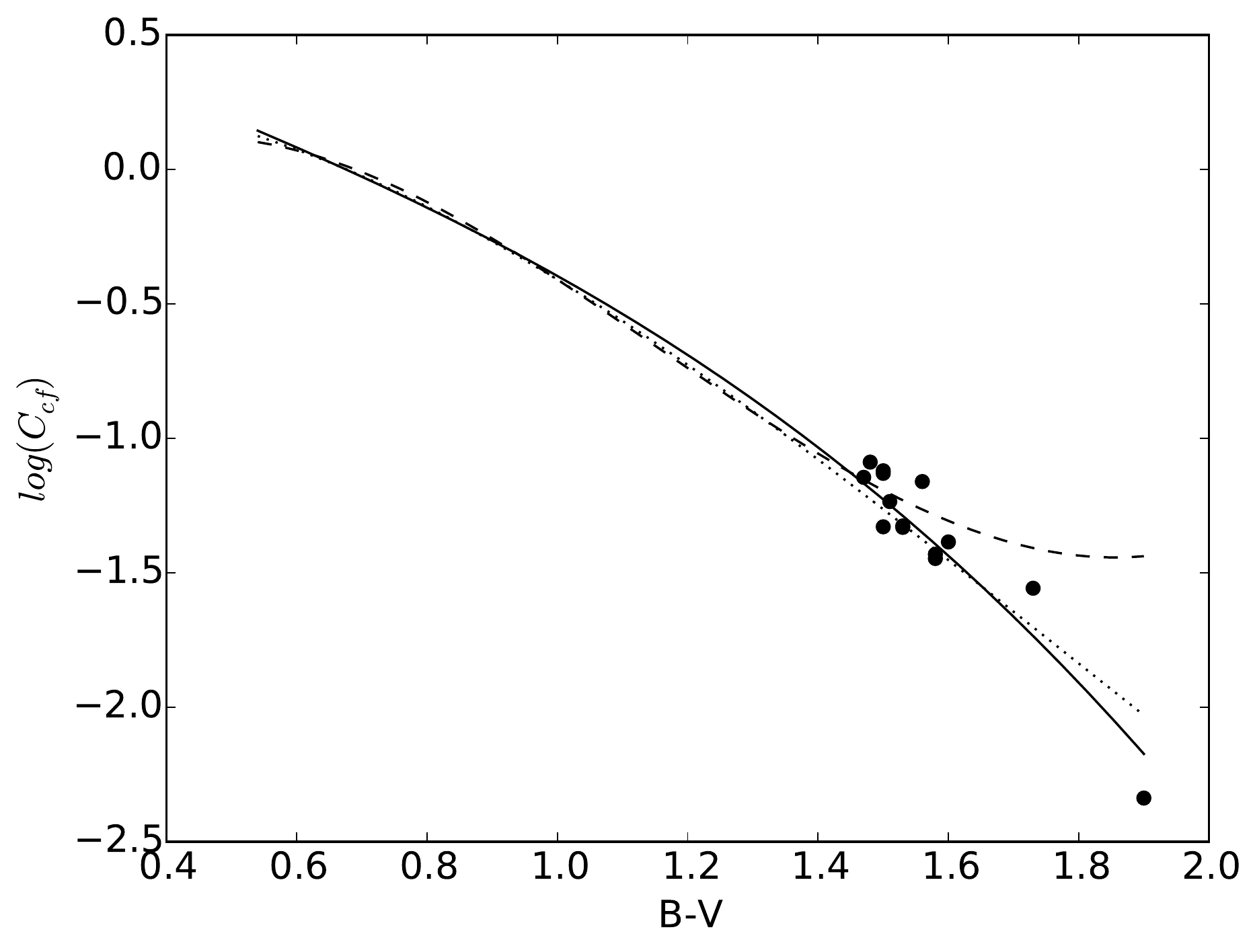}}
\resizebox{\hsize}{!}{\includegraphics{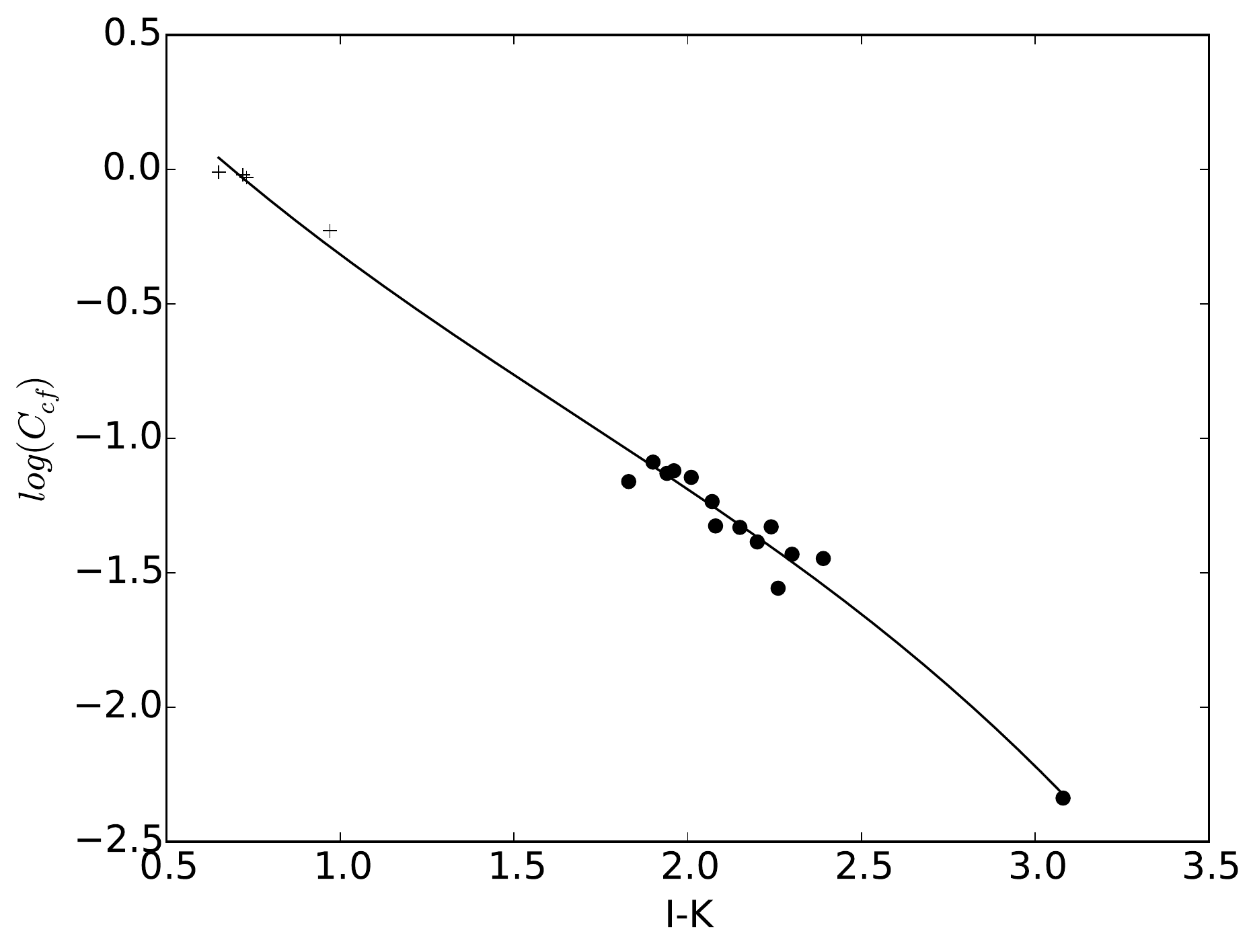}}
\resizebox{\hsize}{!}{\includegraphics{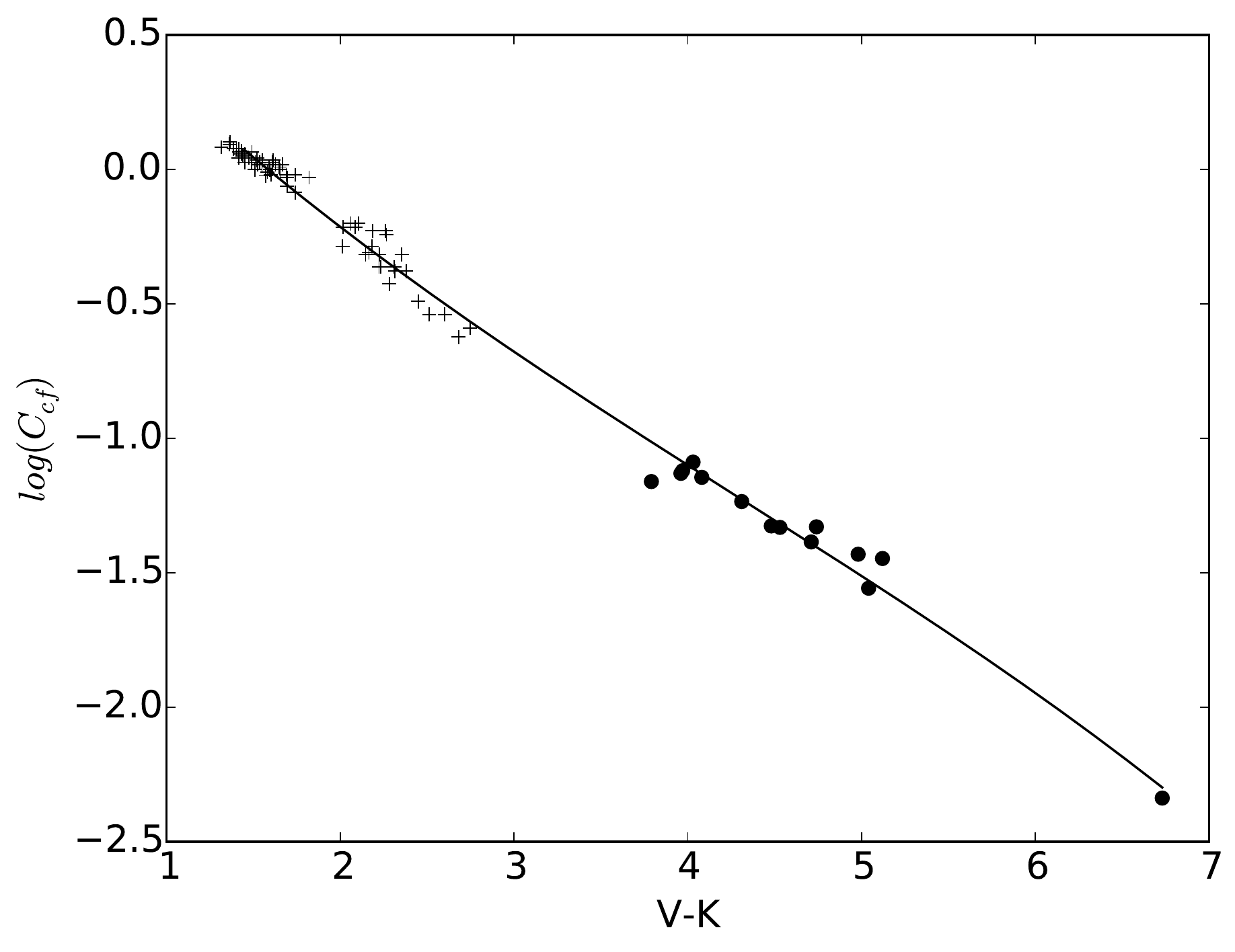}}
\caption{The $C_cf$ bolometric factor against $B-V$ (top panel), $I-K$ 
(middle panel), and $V-K$ (bottom panel). Black dots represent the $C_{cf,\,M}$ 
median, while crosses represent $C_{cf,\,Std}$. Solid lines represent the 
least-square fits of Eq.~(\ref{eq:ccf_sol}) for each color index. 
The top panel omits the 125 $C_{cf,\,Std}(B-V)$ 
data points to more clearly display the \citet{1982A&A...107...31M} 
and \citet{1984A&A...130..353R} fits, represented by dashed and dotted 
lines. The validity ranges of these previous relationships are
$0.45<B-V<1.50$ and $0.3<B-V<1.6,$ respectively, and, for
comparison, 
are extrapolated in this figure up to the color index of M dwarfs.}
\label{fig:ccf_color}
\end{figure}

Since HARPS only observes a small fraction of the spectral energy 
distribution (SED) of our targets, we express $f_{bol}$ as the flux through 
a standard photometric band covered by the HARPS spectra scaled by 
a bolometric correction. We choose the standard visual band and write

\begin{eqnarray}
\label{eq:fbol}
f_{bol} = f_\upsilon \times 10^{-0.4 BC_\upsilon}
,\end{eqnarray}
where $BC_\upsilon$ is the bolometric correction for the visual band, 
which is well determined by \citet{1996ApJ...469..355F} for G-K dwarfs and by
\citet{2001ApJ...548..908L}  for M dwarfs, the two relationships adopted in this work.

Combining Eq.~\ref{eq:fbol} and Eq.~\ref{eq:ccf2}, one obtains:
\begin{eqnarray}
\label{eq:ccf3}
 C_{cf,M}   & = & C_{cf,Std} \times \frac{({f_V+f_R})_M}{({f_V+f_R})_{Std}} 
\frac{f_{\upsilon,Std}}{f_{\upsilon,M}} \times 10^{-0.4(BC_{\upsilon,Std}-BC_{\upsilon,M})}
.\end{eqnarray}

To ensure that the G or K spectra are observed under similar atmospheric 
conditions as the M spectra they are used to pseudo-calibrate, we select 
pairs of M dwarf and G or K spectra that were observed;
\begin{itemize}
\renewcommand{\labelenumi}{\roman{enumi}}
\item within 30 min  
\item at an airmass less than 1.4 for both stars and with an airmass 
  difference of less than 0.05
\item on a good night, as evaluated by a ratio between the measured and 
  synthetic fluxes in the visual band 
  $(\,\Sigma={f_\upsilon / [ t_{exp}10^{-0.4 m_\upsilon} ]}\,)$ 
  such that $\Sigma_{Std} / \Sigma_M -1 \le 0.2.$ 
\end{itemize}

The HARPS M dwarf database contains 14 M dwarfs that fulfill those criteria. 
For each target, we adopt the median to take extra precaution 
allowing us to filter-out eventually imperfectly matched atmospheric conditions 
despite our selection procedure, and protect against the occasional stellar flare. 
Table~\ref{tab:targets_ccf_color_M} lists the resulting median $\log{C_{cf}}$, 
together with the B$-$V, I$-$K, and V$-$K colors. 

Figure~\ref{fig:ccf_color} shows the $C_{cf}$ derived with 
Eq.~(\ref{eq:ccf3}) against B-V, I-K, and V-K, as well as third 
order polynomial least-square fits to these data:

\begin{equation}
\label{eq:ccf_sol}
\qquad \log{C_{cf}}  = c_0\,X^3+c_1\,X^2+c_2\,X+c_3
,\end{equation}

where $X$ is one of the color indexes. Table~\ref{tab:Ccf_Rphot_coef} 
lists the solution for the coefficients of Eq.~(\ref{eq:ccf_sol}), 
the number of data points used for the fit, the rms, and the range of validity 
for each color index. This range of validity corresponds to spectral 
classes G0 to M6. From the rms values, we highlight that the V$-$K and I$-$K relations 
are preferred, but we stress that V$-$K measurements are generally more available
than I$-$K.

\citet{2006MNRAS.372..163J} extrapolated the \citet{1982A&A...107...31M} 
beyond its B-V$<$0.9 stated validity range to compute $R_{HK}$ for M dwarfs. 
The top panel of Fig.~\ref{fig:ccf_color} demonstrates that such an 
extrapolation becomes increasingly invalid for B-V$>$1.5 and  
overestimates $R_{HK}$ by up to a factor of three. Our updated $C_{cf}$-color 
relation matches the latest main sequence stars in \citet{1984A&A...130..353R} 
quite well and allows us to properly compute $R_{HK}$ from S.

\section{The photospheric factor $R_{phot}$}
\label{sec:phot_cor}

\begin{figure}[t]
\centering
\resizebox{\hsize}{!}{\includegraphics{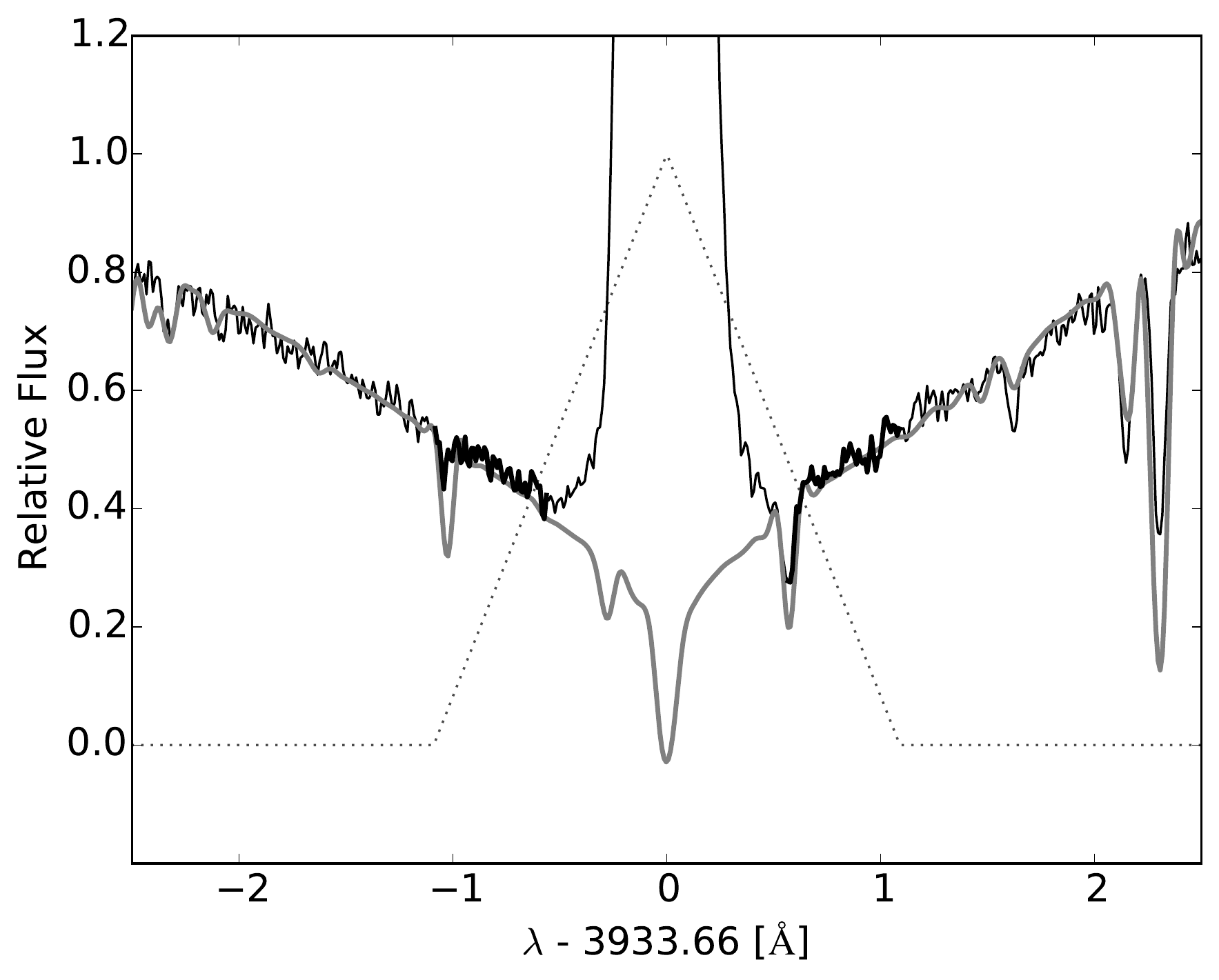}}
\caption{Spectrum of Gl~205 around the Ca \textrm{\small II} K line. The thin 
black and bold gray lines respectively represent the observed spectrum and 
a BT-Settl theoretical spectrum with $T_{eff}=3800$ [K], 
$log(\,g \; [cm/s])=4.5$, and Fe/H = 0.0. The dotted curve depicts the 
K filter. The bold black line represents part of the spectral contribution through 
the triangular K filter used to estimate the photospheric contribution in a previous 
study \citep{1984ApJ...276..254H}.}
\label{fig:replace_synth}
\end{figure}

$R_{HK}$ represents the sum of the photospheric and chromospheric 
fluxes through the two triangular pass-bands  (Fig.~\ref{fig:bands})
of the Mount Wilson system. Separating the chromospheric contribution 
of interest, $R^\prime_{HK}$, therefore requires careful estimation of 
the photospheric flux, $R_{phot}$. 
\citet{1984ApJ...276..254H} and \citet{1984ApJ...279..763N} 
discussed an empirical method for determining $R_{phot}$ for the G-K-dwarfs 
from observed spectra. They use several alternative approaches; all
of them consider the photospheric contribution as the flux outside the
wavelength domain between the two minimum points in the line profiles
of the H and K lines
(see Fig.~\ref{fig:replace_synth} for the K line profile). 
They conclude that arbitrariness in a 
number of choices limits the accuracy of the resulting photospheric 
correction to $\sim$10$\%$. More importantly, \citet{1984ApJ...279..763N} 
pointed out that the photospheric correction becomes unimportant for the 
coolest stars (B-V$\gtrsim$1.0) because the 
reversal emission is very much stronger than the photospheric contribution. 

\begin{figure}[tp]
\centering
\resizebox{\hsize}{!}{\includegraphics{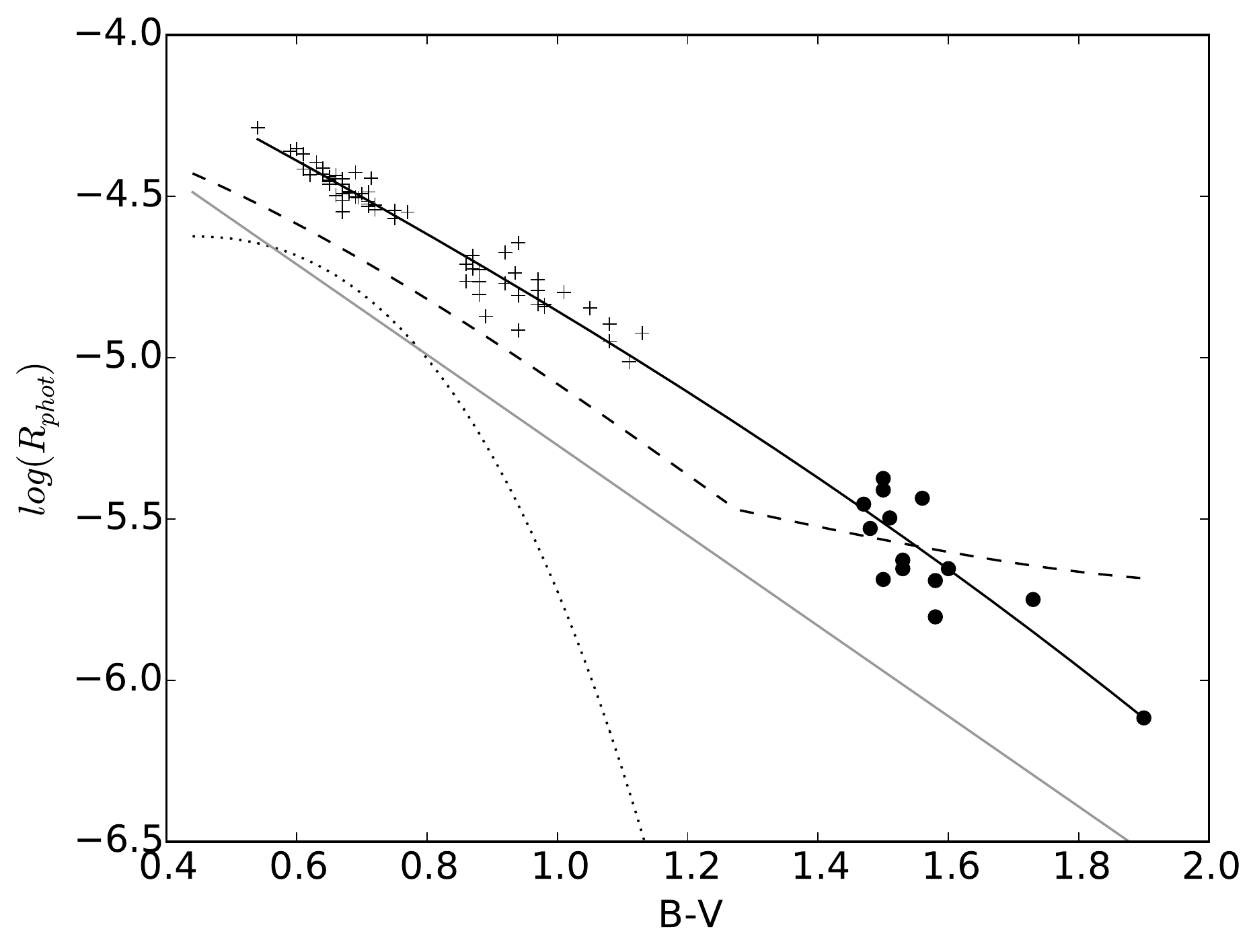}}
\resizebox{\hsize}{!}{\includegraphics{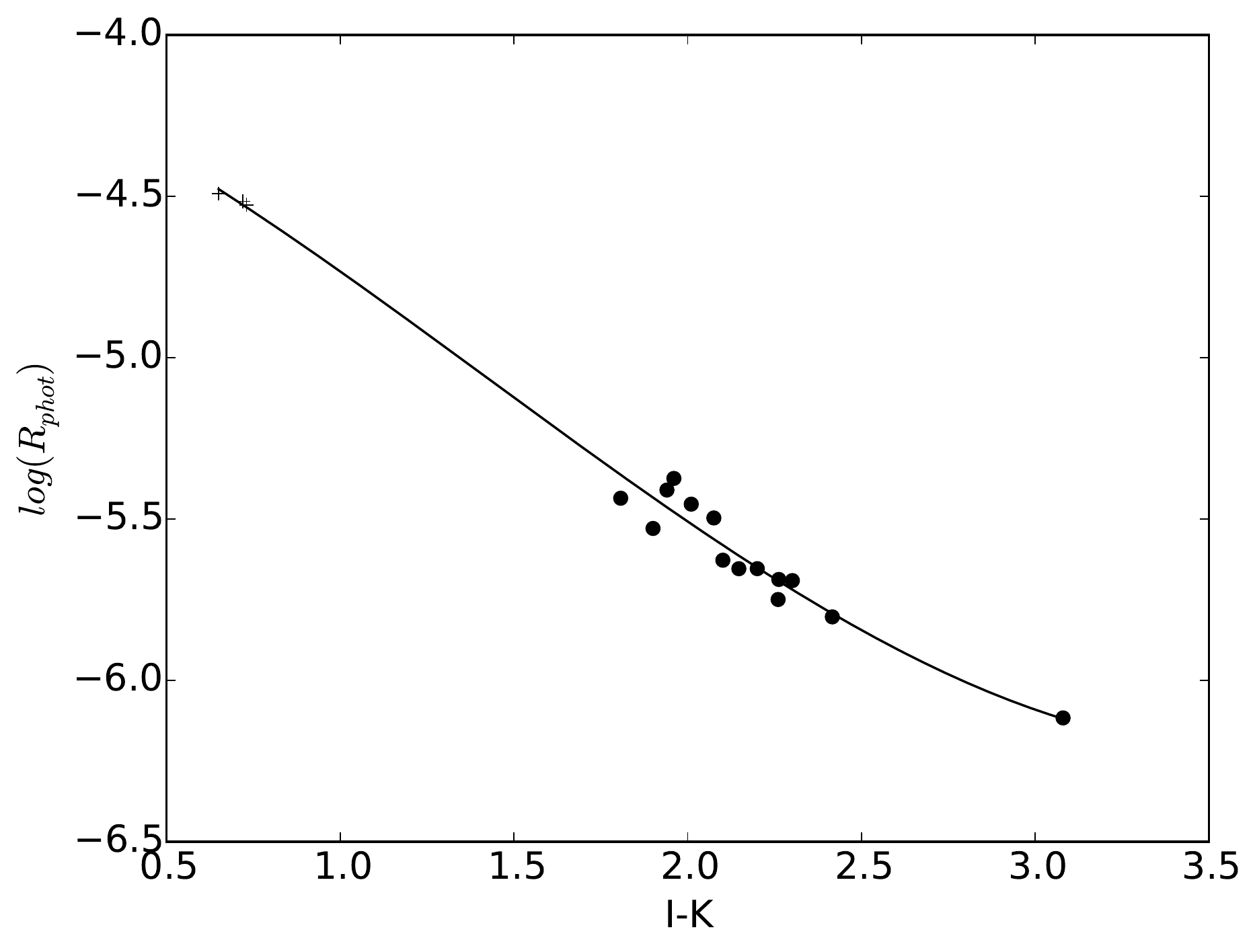}}
\resizebox{\hsize}{!}{\includegraphics{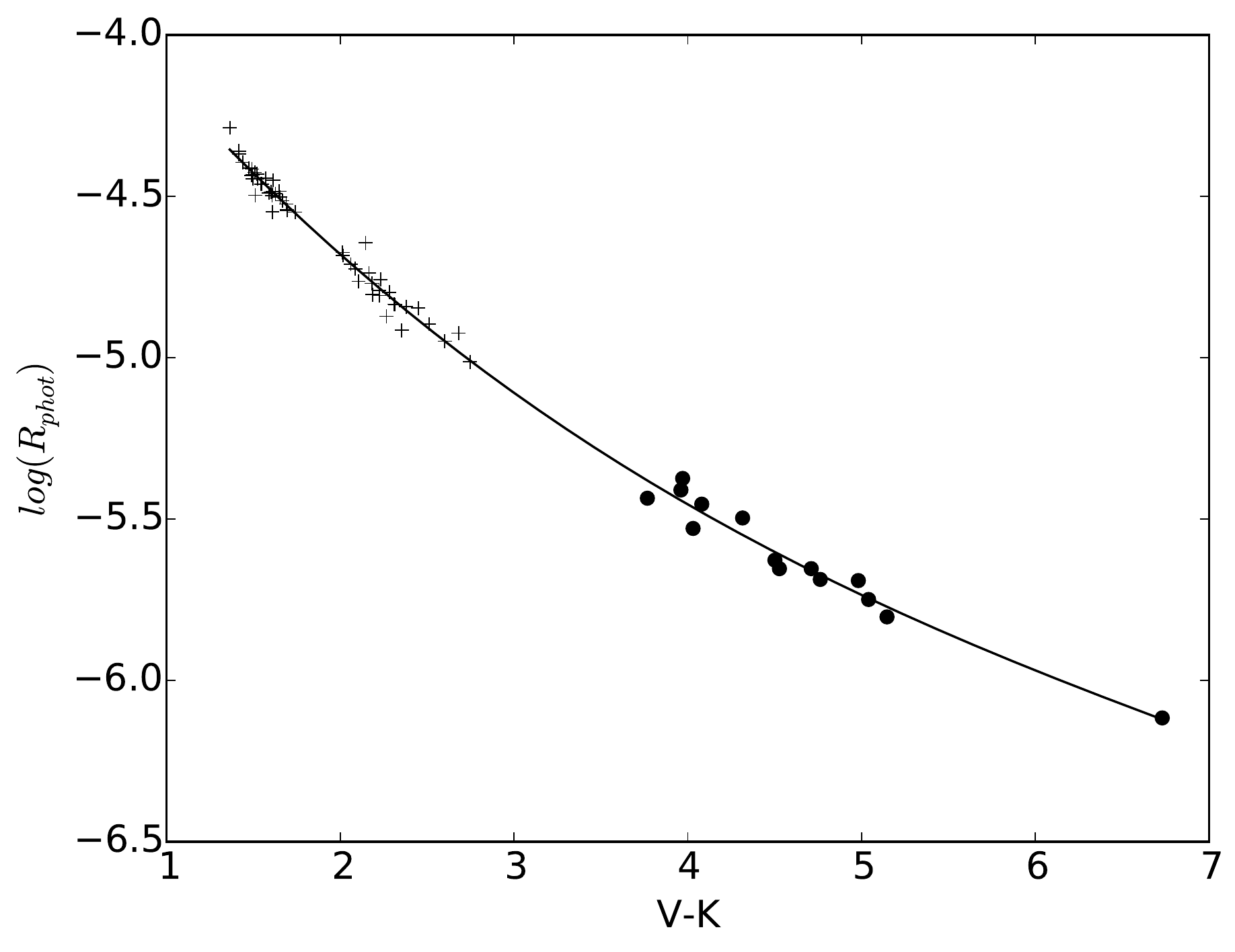}}
\caption{The photospheric factor as a function of, from top to 
bottom, B$-$V, I$-$K, and V$-$K. Dots and crosses 
represent $R_{phot}$ for M and GK-dwarfs, respectively. The solid lines 
are from Eq.~(\ref{eq:rphot_sol}) for each one of the color indexes. 
In the top panel, we compare our relationship for the B$-$V colors
to the ones of \citet{1984ApJ...276..254H}, \citet{1984ApJ...279..763N}, and 
\citet[][only from synthetic spectra]{2013A&A...549A.117M}  
in dotted, dashed, and dash-dotted lines, respectively.
Their 
range of validity is $0.44<B-V<0.82$ for \citet{1984ApJ...276..254H} and 
\citet{1984ApJ...279..763N}, and $0.44<B-V<1.6$ for \citet{2013A&A...549A.117M}.
}
\label{fig:rphot_color}
\end{figure}

Like previous authors \citep[e.g.,][]{2009AJ....137.3297W,2013A&A...549A.117M}, we therefore elect to use a theoretical grid of photospheric 
model, \citep[here BT-Settl/CIFIST2011bc][]{IAU:9137085}\footnote{http://phoenix.ens-lyon.fr/Grids/} 
to evaluate the photospheric contribution. Figure~\ref{fig:replace_synth} 
illustrates an example of the match between one of these models and the 
observed spectra of an early-M dwarf around the \ion{Ca}{ii} K line.

For each target, we compute $\log{g}$ from the mass and radius
that we obtain from the relations of 
\citet{2000A&A...364..217D} and \citet{ 2012ApJ...757..112B}, respectively. We obtain
T$_{\rm eff}$ from \citet{2012ApJ...757..112B} and [Fe/H] from 
\citet{2013A&A...551A..36N}. We then inspect the model spectra for 
grid points within T$_{eff}\pm200$ [K] and $log(g)\pm0.5$~[$cm\,s^{-1}$] 
and select that which visually best matches the average observed 
spectrum of the star between 3880 and 4022 $\AA$. 
Table~\ref{tab:Rphot_BT-Settl} lists our computed stellar parameters 
as well as those of the model that best matches the average
spectrum of each star.

We then return to the individual observed spectra and normalize the 
model spectrum to match them over the wings of (separately) the 
Ca \textrm{\small II} H and K lines, which are free 
of chromospheric emission (Fig.~\ref{fig:replace_synth}). We then 
replace the central 2~$\AA$ region of each \ion{Ca}{ii} 
H and K line with the normalized synthetic spectrum 
(Fig.~\ref{fig:replace_synth}), to obtain a hybrid spectrum that
contains no chromospheric emission, from which we compute  $S_{phot}$ 
using Eq.~(\ref{eq:s_harps}) and (\ref{eq:s_harps_wright_lin}).
Using Eq.~(\ref{eq:ccf_sol}) for V$-$K,  
we obtain the photospheric contribution $R_{phot}$ in 
Eq~(\ref{eq:Rhk}) as:

\begin{equation}
\label{eq:Rphot}
R_{phot} = K \cdot \sigma^{-1} \cdot 10^{-14} \cdot C_{cf} \cdot S_{phot}
.\end{equation}

As explained in Sect.~\ref{sec:scaling_S}: $K \cdot \sigma^{-1} \cdot 10^{-14}$
=1.887$\times$10$^{-4}$. 
Table~\ref{tab:targets_ccf_color_M} lists the median of 
the individual $R_{phot}$ measurements for each M dwarf 
and Table~\ref{tab:targets_ccf_color_std} gives this value for our GK-dwarf
calibrators.

Figure~\ref{fig:rphot_color} shows the median $R_{phot}$ obtained against 
B$-$V, I$-$K, and V$-$K, as well as least-square third order polynomial fits:

\begin{equation}
\label{eq:rphot_sol}
\qquad log(R_{phot})  = r_0\,X^3+r_1\,X^2+r_2\,X+r_3
,\end{equation}
where $X$ stands for one of the color indexes. The values for the 
coefficients are tabulated in Table~\ref{tab:Ccf_Rphot_coef}, 
in the same format as described in Sect.~\ref{sec:bol_cor} for 
the bolometric factor $C_{cf}$. The $R_{phot}$  vs V$-$K relationship
shows the lowest dispersion.
\begin{table}[t]
\centering
\caption{Solutions for coefficients for the $\log{C_{cf}}$-color and $R_{phot}$-color relationships.}
\label{tab:Ccf_Rphot_coef}
\begin{tabular*}{\hsize}{@{\extracolsep{\fill}}lrrr}

\hline \hline
\noalign{\smallskip}
\multicolumn{4}{l}{$\log{C_{cf}}$} \\

Coefficient & \multicolumn{1}{c}{$B-V$} & \multicolumn{1}{c}{$I-K$} & \multicolumn{1}{c}{$V-K$}\\

\noalign{\smallskip}
\hline
\noalign{\smallskip}

$c_0$ & $-0.203\pm0.008$ & $-0.082\pm0.005$ & $-0.005\pm0.000$ \\
$c_1$ & $0.109\pm0.088$  & $0.416\pm0.180$  & $0.071\pm0.000$ \\
$c_2$ & $-0.972\pm0.099$ & $-1.544\pm0.534$ & $-0.713\pm0.006$ \\
$c_3$ & $0.669\pm0.011$  & $0.894\pm0.120$  & $0.973\pm0.006$ \\

\noalign{\smallskip}
\hline
\noalign{\smallskip}

N data points & \multicolumn{1}{c}{140} & \multicolumn{1}{c}{18} & \multicolumn{1}{c}{81} \\
RMS $\log{C_{cf}}$ & \multicolumn{1}{c}{0.102} & \multicolumn{1}{c}{0.064} & \multicolumn{1}{c}{0.062} \\
RMS $C_{cf}$ & \multicolumn{1}{c}{0.0111} & \multicolumn{1}{c}{0.0078} & \multicolumn{1}{c}{0.0088} \\
Valid range & \multicolumn{1}{c}{0.54 -- 1.9} & \multicolumn{1}{c}{0.72 -- 3.08} & \multicolumn{1}{c}{1.45 -- 6.73} \\

\noalign{\smallskip}
\hline \hline
\noalign{\smallskip}

\multicolumn{4}{l}{$R_{phot}$} \\

Coefficient & \multicolumn{1}{c}{$B-V$} & \multicolumn{1}{c}{$I-K$} & \multicolumn{1}{c}{$V-K$}\\

\noalign{\smallskip}
\hline
\noalign{\smallskip}

$r_0$ & $-0.045\pm0.033$ & $0.056\pm0.006$  & $-0.003\pm0.000$ \\
$r_1$ & $-0.026\pm0.392$ & $-0.237\pm0.188$ & $0.069\pm0.000$ \\
$r_2$ & $-1.036\pm0.470$ & $-0.453\pm0.547$ & $-0.717\pm0.003$ \\
$r_3$ & $-3.749\pm0.056$ & $-4.099\pm0.115$ & $-3.498\pm0.004$ \\

\noalign{\smallskip}
\hline
\noalign{\smallskip}

$R_{phot}$ data points & \multicolumn{1}{c}{78} & \multicolumn{1}{c}{17} & \multicolumn{1}{c}{67} \\
$log(R_{phot})$ RMS & \multicolumn{1}{c}{0.104} & \multicolumn{1}{c}{0.056} & \multicolumn{1}{c}{0.040} \\
$R_{phot}$ RMS & \multicolumn{1}{c}{$6.5\times 10^{-7}$} & \multicolumn{1}{c}{$4.4\times 10^{-7}$} & \multicolumn{1}{c}{$3.0\times 10^{-7}$} \\
\noalign{\smallskip}
Valid range & \multicolumn{1}{c}{0.54 -- 1.9} & \multicolumn{1}{c}{0.65 -- 3.08} & \multicolumn{1}{c}{1.36 -- 6.73} \\

\noalign{\smallskip}
\hline
\noalign{\smallskip}

\end{tabular*}
\end{table}
\begin{table}[t]
\centering
\caption{Measured stellar parameters, and stellar parameters for the 
BT-Settl model, which best matches the observed spectrum 
around the Ca \textrm{\small II} H and K lines. 
}
\label{tab:Rphot_BT-Settl}
\begin{tabular*}{\hsize}{@{\extracolsep{\fill}}lccc}

\hline
\noalign{\smallskip}

Name & log(g [cm s$^{-1}$]) & T$_{\rm eff}$ [K]& Fe/H\\
& Det. / Used & Det. / Used & Det. / Used\\

\noalign{\smallskip}
\hline
\noalign{\smallskip}

Gl~1   & 5.00 / 5.0 & 3458 / 3500 & -0.45 / -0.5 \\
Gl~191 & 5.25 / 4.5 & 3134 / 3300 & -0.88 / -1.0 \\
Gl~205 & 4.70 / 4.5 & 3780 / 3800 & +0.22 / -0.0 \\
Gl~229 & 4.82 / 5.0 & 3643 / 3700 & -0.10 / -0.0 \\
Gl~393 & 4.75 / 5.0 & 3639 / 3600 & -0.22 / -0.0 \\
Gl~551 & 4.94 / 5.0 & 2659 / 2800 & -0.00 / -0.0 \\
Gl~581 & 4.92 / 5.0 & 3327 / 3500 & -0.21 / -0.0 \\
Gl~588 & 4.70 / 4.5 & 3519 / 3500 & +0.07 / -0.0 \\
Gl~628 & 4.80 / 5.0 & 3364 / 3400 & -0.02 / -0.0 \\
Gl~674 & 4.91 / 5.0 & 3374 / 3400 & -0.25 / -0.5 \\
Gl~699 & 5.30 / 5.0 & 3088 / 3300 & -0.52 / -0.5 \\
Gl~849 & 4.59 / 4.5 & 3519 / 3500 & +0.24 / -0.0 \\
Gl~876 & 4.69 / 4.5 & 3421 / 3400 & +0.15 / -0.0 \\
Gl~887 & 4.89 / 5.0 & 3686 / 3700 & -0.24 / -0.0 \\

\noalign{\smallskip}
\hline
\noalign{\smallskip}
\end{tabular*}
\end{table}

We note a systematic offset between our measurements and 
the \citet{1984ApJ...276..254H} or \citet{1984ApJ...279..763N} 
$R_{phot}$ fits, with their  $R_{phot}$ being systematically lower. 
We suspect that the offset reflects low-level errors in the 
extrapolation that is intrinsic to a purely empirical estimation 
of the photospheric flux under a chromospheric line, but it
could, in principle, instead reflect systematics in the theoretical
spectra that we use. We note, however, that our approach 
is more correct from a physical point of view.

\begin{figure*}[t]
\centering
\resizebox{\hsize}{!}{\includegraphics{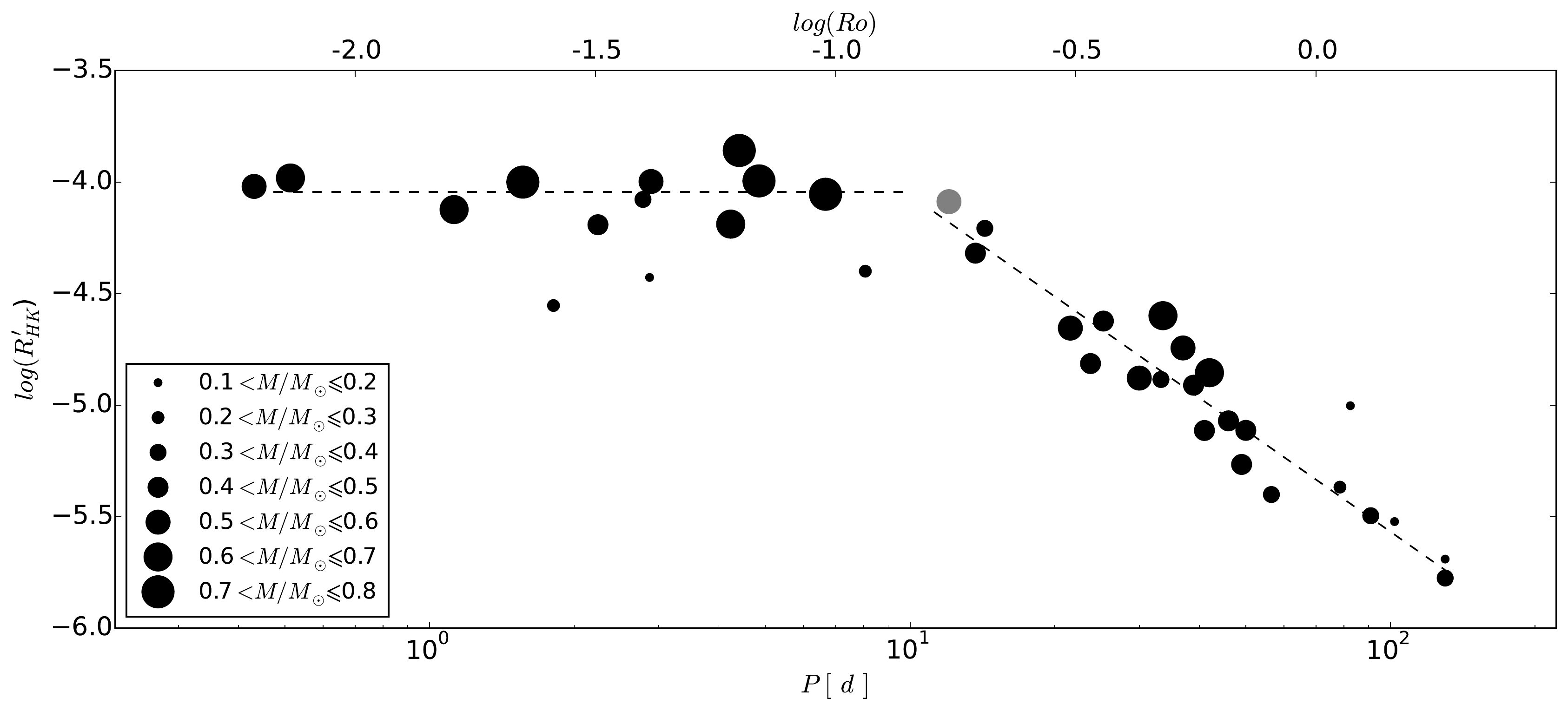}}
\caption{$log(R^\prime_{HK})$ against rotation period. The area of the 
filled circles is proportional to the stellar mass, with the gray circle
representing one star that only has a photometric parallax 
\citep{2011AJ....142..138L} and consequently a poorly determined mass.
$log(R^\prime_{HK})$ saturates for $P_{rot} < 10$~days, and then decreases 
as an approximately linear function of  $log(P_{rot})$. 
For illustration, the upper x-axis displays the Rossby number 
($Ro=P_{rot}/\tau_c$) for an assumed $\tau_c=70$ d.
}
\label{fig:logR_Prot}
\end{figure*}

\section{Accuracy of the $R^\prime_{HK}$ calibration}
\label{sec:Rhk_accuracy}

The $R^\prime_{HK}$ comes from, as expressed in Eq.~(\ref{eq:Rhk}),  the product 
of a constant $(K \cdot \sigma^{-1} \cdot 10^{-14})$ and of two factors: $C_{cf}$ and 
$(S-S_{phot})$. Accuracy and precision of the $R^\prime_{HK}$ calibration is then 
directly connected to both.

\subsection{The $\log{C_{cf}}$ accuracy}

Our updated $\log{C_{cf}}$ relationship, that includes M dwarfs, is purely 
empirically determined as the product of the $C_{cf}$ of G-K dwarf calibrators 
of the flux ratio between M dwarfs and G-K dwarf calibrators in the two-wavelength 
domain of our HARPS spectra (V and R calcium pseudo-continuum and standard visual 
bands), and of a bolometric correction. The internal precision of $\log{C_{cf}}$ 
is given in Table~\ref{tab:Ccf_Rphot_coef}, with a dispersion of approximately 0.06 dex. 
We discuss here the source of possible systematic errors generating bias. 

The $\log{C_{cf}}$ of G-K dwarf calibrators is computed with the \citet{1982A&A...107...31M} 
relationship whose origins are the Mount Wilson measurements in the 1960s and 
1970s. It is probable that such a factor is too poorly flux calibrated 
\citep[see the discussion of][]{2013A&A...549A.117M} 
to be considered as an accurate bolometric factor for the V and R calcium pseudo-continuum 
band. However, that does not impact our study since our goal is to obtain $R^\prime_{HK}$ 
for M dwarf that can be compared to the ones of G-K dwarfs. Therefore if we define our 
$C_{cf}$ relationship as a calibration anchored on its equivalent for G-K dwarfs, 
this step does not create a bias.

We compute the ratio of fluxes
\footnote{$C_{cf} \propto [({f_V+f_R})_M/({f_V+f_R})_{Std}]\ / \ 
[f_{\upsilon,Std}/f_{\upsilon,M}]$; from Eq.~(\ref{eq:ccf3}).} 
passing through different bands when deriving the bolometric factor. 
As described in Sect.~\ref{sec:bol_cor}, we ensure that the G-K standards 
are observed under similar atmospheric conditions as the M dwarfs. Thus, 
if the signal-to-noise of spectra is sufficient, this step can 
only contribute to the dispersion of the $C_{cf}$. However, under low 
signal-to-noise conditions, an imperfect background subtraction can occur 
for HARPS reduced spectra and may generate an instrumental systematic error. 
This effect may impact the flux measurements in the V and R calcium 
pseudo-continuum band and then our $C_{cf}$ calibration. This could be particularly 
significant for M dwarfs, due to their low flux at the blue wavelengths 
of the \ion{Ca}{ii} H and K lines. \citet{2011arXiv1107.5325L} examine systematic errors 
in the HARPS spectrum in this wavelength domain and found that $R^\prime_{HK}$ 
measurements from HARPS spectra hit an instrumental  noise floor only when 
the S-index photon uncertainty becomes smaller than $\sim$0.007 dex.
To confirm this behavior, for M dwarfs, we plotted   the 
$R^\prime_{HK}$ standard deviation of our sample normalized to the $R^\prime_{HK}$ mean value 
against the average of the S/N ratio (per pixel) in the V and R calcium 
pseudo-continuum bands (S/N$_{VR,\,pix}$), and observe that its lower envelope 
rises systematically for S/N$_{VR,\,pix}~\lesssim$5. We attribute this rise 
to systematic error, including imperfect background subtraction. To ensure that 
our $C_{cf}$ calibration is not affected by this effect, we only retain spectra 
with S/N$_{VR,\,pix}\geq$5. The average signal-to-noise of the 14 M dwarfs and of 
their comparison G-K-dwarfs that we use is approximately 30 
(ranging from 7.5 to 57.6). At such a level of flux, the effect of the background 
subtraction is negligible.

$R^\prime_{HK}$ being, in theory, the ratio between the chromospheric emission 
in the Ca H and K lines and the bolometric flux, we have to use a bolometric correction. 
For that, we use two different bolometric correction 
relationships: \citet{1996ApJ...469..355F} for G-K dwarf calibrators and 
\citet{2001ApJ...548..908L} for M dwarfs. 
Both relations are obtained from flux distribution measurements that are 
completed by synthetic spectra when certain wavelength domains are uncovered. 
They use similar zero points (respectively $M_{bol, \odot}$ = 4.73 and 4.75). 
As demonstrated  by \citet{1996ApJ...469..355F, 2010AJ....140.1158T} 
for G-K dwarfs and by \citet{2014PASP..126..642S, 2015ApJ...804...64M} 
for M dwarfs, for example, the agreement among different bolometric corrections is usually 
better than 0.1~mag. We therefore do not expect that this step introduces 
systematic error higher that 0.1~mag on $\log{C_{cf}}$.

To conclude, we are confident that systematic error in our $\log{C_{cf}}$ 
for M dwarfs is negligible.

\subsection{Accuracy on $(S-S_{phot})$}

It is more difficult to estimate the systematic error on $S_{phot}$, since we use 
synthetic spectra to reproduce the core of the photospheric \ion{Ca} {ii} H and K 
absorption lines. Such line cores cannot be measured directly, being always hidden 
by the chromospheric emission. If the synthetic model does not reproduce them correctly, 
it will introduce a bias in our $R_{phot}$ relationship. The $\log{R_{phot}}$ has a 
dispersion of approximately 0.05 dex (Table~\ref{tab:Ccf_Rphot_coef}).

We make sure that the wings of photospheric \ion{Ca} {ii} H and K lines are well 
reproduced by using theoretical spectra (see Fig~\ref{fig:replace_synth}). If, 
however, the core of the photospheric line is not as well reproduced as the wings, 
we point out that this part of the photospheric contribution covers approximately 25\% of the 
wavelength domain of the triangular spectrum used to measure $S$. The model of 
the photospheric flux used
is intermediate between two extreme solutions: assuming that 
photospheric flux is 0 under the chromospheric line or that it is constant and 
at the level of the reversal points at the basis of the emission line. Therefore,
our estimation will not differ by more than $\sim$15\% from these two extreme 
solutions. \citet{1984ApJ...276..254H} come to the same conclusion on the effect of 
different methods to estimate $S_{phot}$. We stress that the use of theoretical spectra 
is today the most physically motivated approach to address this issue.

Furthermore, the factor involved in the error budget is $(S-S_{phot})$. In our study 
the quietest (with lowest \ion{Ca} {ii} H and K emission) M dwarf used in the calibration 
has a $S$ 42\% higher than $S_{phot}$, minimizing the impact of the systematic error on $S_{phot}$ 
to less than 10\%. Taking everything into account, we estimate that the effect of such 
systematic error is below 0.1~mag on $\log{R^\prime_{HK}}$.

\section{$R^\prime_{HK}$ versus rotation}
\label{sec:Rhk_rotation}

For low-mass stars the magnetic field is generated by a combination
of (i) $\alpha\Omega$ dynamo \citep[e.g.,][]{1955ApJ...122..293P} taking place 
under the presence of a radiative core separated from a convective envelope 
by a strongly sheared thin layer (the tachocline), and (ii) a fully convective 
dynamo \citep[later than M4V; e.g.,][]{2005ApJ...631..529B} showing similarities 
with a planetary one \citep{2013A&A...549L...5G, 2014A&A...564A..78S}. Stars later 
than M4V become fully convective \citep{2005ApJ...631..529B} and only the second 
type of dynamo is at work.

On the one hand, Zeeman Doppler Imaging of active 
M dwarfs indicates that the topology of the large-scale component of their 
magnetic field changes when they approach the full-convection limit 
\citep{2008MNRAS.390..545D,2008MNRAS.390..567M}. On the other, no obvious 
change at the fully-convective transition is noticeable in the relationship 
between the rotational period and the average field strength 
\citep{2007ApJ...656.1121R, 2009ApJ...692..538R}, or its proxies 
such as magnetic activity \citep[e.g., $L_X$][]{2007AcA....57..149K}.

Here we analyze how $R^\prime_{HK}$ varies with stellar rotation around the 
fully convective limit. We use 38 M1 to M6 dwarfs with known stellar rotation 
periods, inferred from either photometric periodicities or periodic modulation 
of the S-index. The majority of periods are obtained from the literature 
(Table~\ref{tab:Prot_RHK}), but those for 7 M dwarfs are inferred from strong peaks 
(above 0.3\%FAP) in the respective periodogram \citep[e.g.,][]{2009A&A...496..577Z} 
of our measurements of its \ion{Ca}{ii} H and K emission (GJ~3138, Gl~654, Gl~752A, 
Gl~876, Gl~880, Gl~382, Gl~514). Both $C_{cf}$ and $R_{phot}$ correlate with a similar 
level of dispersion with either $V-K$ or $I-K$ (Table~\ref{tab:Ccf_Rphot_coef}). The 
dispersion is superior in the correlation with $B-V$. 
To compute $R^\prime_{HK}$  , we use, hereafter, $\log{C_{cf}}(V-K)$ and $\log{R_{phot}}(V-K)$ 
relationships, because, in general, $V-K$ photometry is more available than $I-K$.

For solar type stars, $log(R^\prime_{HK})$ correlates better with the Rossby 
number ($Ro=P_{obs}/\tau_c$, where $\tau_c$ is the convective overturn 
time) than with $P_{rot}$ alone \citep{1984ApJ...279..763N}. This matches the 
theoretical expectation that the strength of an $\alpha\Omega$ 
dynamo process is proportional to $Ro^{-2}$. $Ro$ is therefore
widely used when relating magnetic activity with rotation, though
some authors argue that $P_{rot}$ should be prioritized 
\citep{1993IAUS..157..141S,2014ApJ...794..144R}. $\tau_c$ 
can be determined either empirically \citep[e.g.,][]{1984ApJ...279..763N} 
or theoretically \citep[e.g.,][]{1998A&A...334..953V}, with both approaches
giving uncertain results for M dwarfs. 

Fig.~\ref{fig:logR_Prot} demonstrates that $log(P_{rot})$ correlates closely 
with $log(R^\prime_{HK})$ for stellar masses between 0.1M$_\odot$ and 0.8M$_\odot$.
Above a $\sim$10-day rotation period, activity decreases with slower rotation,
while below that period it no longer depends on rotation. The 
$log(R^\prime_{HK})$-$log(P_{rot})$ relationship is well described by:

\begin{equation}
\label{eq:R_Prot_relation}
log(R^\prime_{HK}) = \left\{ \begin{array}{ll}
  -1.509 \cdot log(P_{rot})-2.550 &\mbox{ if $P_{rot}[d]>10$} \\
  -4.045 &\mbox{ if $P_{rot}[d]<10$}
       \end{array} \right.
,\end{equation}
where the $log(R^\prime_{HK})$ average for the saturated regime does not account 
for the three stars with masses below 0.2M$_\odot$ (GJ~3379, Gl~729, G141-29; see 
justification below), and Gl~551 is not considered for the non-saturated fit as it shows flares 
with a significantly higher cadence than their siblings with approximately the same age or rotational 
period \citep[e.g.,][]{2016ApJ...829L..31D}.
The $log(R^\prime_{HK})$ uncertainty is $\pm0.093$ if $P_{rot}[d]<10$. 
For $P_{rot}[d]>10$, the slope and the y-intercept uncertainties from the fit 
are $\pm0.007$ and $\pm0.020$, respectively. The rms of the non-saturated regime 
in Eq.~(\ref{eq:R_Prot_relation}) is 7.68 days while the median of the relative uncertainty 
is 8.7\%.

Activity saturation in fast M dwarf rotators is a well known behavior 
observed in many proxies $ L_{H\alpha}/L_{bol}$, $L_X/L_{bol}$ , or $B_f$ 
\citep[e.g.,][]{1998A&A...331..581D,2007AcA....57..149K,2009ApJ...692..538R,
2012AJ....143...93R}, and is interpreted either as a physical saturation of 
the dynamo process  or as active regions completely covering the stellar 
surface while the magnetic field continues to grow. For M dwarfs, $R^\prime_{HK}$ 
saturates at $Ro \approx 0.1$ \citep[adopting here 
$\tau_c=70\ d$, following][]{2009ApJ...692..538R}. 
$L_X/L_{bol}$ and $Bf$ similarly saturate for $Ro \approx 0.1$ 
\citep{2007AcA....57..149K,2009ApJ...692..538R}. 
Above 10~days and up to $\sim$100 days, $log(R^\prime_{HK})$ varies 
linearly with $log(P_{rot})$, as also observed for FGK-dwarfs 
\citep{1984ApJ...279..763N, 2008ApJ...687.1264M}.

11 of our 38 M dwarfs have masses under the 0.35M$_\odot$ full convection 
limit. They therefore have no tachocline and consequently cannot operate 
an $\alpha\Omega$ dynamo. Fig.~\ref{fig:logR_Prot} demonstrates that the 
$log(R^\prime_{HK})$ vs. $log(P_{rot})$ relationship does not markedly change at 
this transition, and that stellar rotation continue to drive activity in 
fully convective stars. We note that the three lowest-mass stars in the 
saturated half of Fig.~\ref{fig:logR_Prot} have lower $log(R^\prime_{HK})$ 
than their more massive counterparts. 
These three M dwarfs were not considered for the saturated fit 
as we suspect that they may follow a different regime. However, a larger sample 
will be needed to confirm if the $log(R^\prime_{HK})$ level in the saturated 
regime is correlated with stellar mass for mid- to late-M dwarfs.

\begin{table}
\centering
\caption{The $log{R^\prime_{HK}}$ (fourth column) and rotation periods (fifth column) of the 38 
M dwarfs for which both are known. Their references are given in the sixth column: 
(1) \citet{2007AcA....57..149K}, (2) \citet{2013A&A...549A.109B}, 
(3) \citet{2008MNRAS.390..567M}, (4) \citet{2011ApJ...727...56I}, 
(5) \citet{2014Sci...345..440R}, (6) \citet{2015A&A...575A.119A}, 
and (7) refer to this work. The seventh column gives the rotation periods derived from 
Eq.~(\ref{eq:R_Prot_relation}), where objects flagged with $^{(*)}$ are the three 
very-low-mass stars showing the lowest log($R^\prime_{HK}$) in the saturated regime 
(Fig.~\ref{fig:logR_Prot}).  The rms of P$_{Rot.}$ and P$_{Rot.\ Fit}$ 
is 12 days. V-K color and stellar mass are tabulated 
in the second and third columns,  respectively.}
\label{tab:Prot_RHK}
\begin{tabular*}{\hsize}{@{\extracolsep{\fill}}lcclccc}

\hline
\noalign{\smallskip}

Name & V-K & M & log($R^\prime_{HK}$) & P$_{Rot.}$ & P$_{Ref.}$ & P$_{Rot.\ Fit}$ \\
& [mag] & [M$_\odot$]  & & [d] & & [d]\\

\noalign{\smallskip}
\hline
\noalign{\smallskip}

GJ~1264  &      4.288   &       0.74  & -4.055  &       6.67    & (1)& $\leq$10\\
Gl~699   &      5.040   &       0.16  & -5.691  &       130     & (1)& 121\\
Gl~569A  &      4.416   &       0.48  & -4.319  &       13.68   & (1)& 15\\
GJ~182   &      3.680   &       0.79  & -3.859  &       4.41    & (1)& $\leq$10\\
GJ~890   &      3.852   &       0.57  & -4.020  &       0.43    & (1)& $\leq$10\\
GJ~867A  &      4.716   &       0.63  & -4.189  &       4.23    & (1)& 12\\
GJ~841A  &      4.769   &       0.68  & -4.124  &       1.12    & (1)& $\leq$10\\
Gl~803   &      4.230   &       0.74  & -3.995  &       4.85    & (1)& $\leq$10\\
Gl~729   &      5.080   &       0.17  & -4.428  &       2.87    & (1)& 18 $^{(*)}$\\
Gl~618A  &      4.686   &       0.38  & -5.401  &       56.52   & (1)& 78\\
Gl~551   &      6.730   &       0.12  & -5.003  &       82.53   & (1)& 42\\
GJ~494   &      4.150   &       0.60  & -3.998  &       2.89    & (1)& $\leq$10\\
GJ~431   &      4.980   &       0.37  & -4.208  &       14.31   & (1)& 13\\
GJ~3367  &      3.817   &       0.54  & -4.088  &       12.05   & (1)& $\leq$10\\
GJ~1054A &      3.949   &       0.66  & -3.982  &       0.51    & (1)& $\leq$10\\
GJ~103   &      3.941   &       0.75  & -4.000  &       1.56    & (1)& $\leq$10\\
Gl~205   &      4.080   &       0.63  & -4.599  &       33.61   & (1)& 23\\
Gl~358   &      4.660   &       0.42  & -4.623  &       25.26   & (1)& 24\\
Gl~176   &      4.509   &       0.49  & -4.911  &       38.92   & (1)& 37\\
Gl~674   &      4.480   &       0.34  & -4.885  &       33.29   & (1)& 35\\
Gl~479   &      4.640   &       0.43  & -4.814  &       23.75   & (2)& 32\\
Gl~526   &      4.010   &       0.49  & -5.113  &       50.00   & (2)& 50\\
Gl~388   &      4.710   &       0.42  & -4.191  &       2.24    & (3)& $\leq$10\\
Gl~12    &      4.809   &       0.22  & -5.368  &       78.50   & (4)& 74\\
G~141-29 &      5.235   &       0.24  & -4.400  &       8.07    & (4)& 17 $^{(*)}$\\
GJ~3379  &      5.334   &       0.23  & -4.554  &       1.81    & (4)& 21 $^{(*)}$\\
GJ~1057  &      5.950   &       0.18  & -5.522  &       102.00  & (4)& 93\\
LHS~1610 &      5.783   &       0.17  & -5.375  &       78.80   & (4)& 75\\
Gl~285   &      5.420   &       0.31  & -4.078  &       2.78    & (4)& $\leq$10\\
Gl~581   &      4.710   &       0.31  & -5.776  &       130     & (5)& 137\\
GJ~3293  &      4.520   &       0.52  & -5.114  &       41      & (6)& 50\\
GJ~3138  &      3.710   &       0.68  & -4.855  &       42      & (7)& 34\\
Gl~654   &      4.120   &       0.48  & -5.266  &       49      & (7)& 63\\
Gl~752A  &      4.460   &       0.49  & -5.071  &       46      & (7)& 47\\
Gl~876   &      5.120   &       0.33  & -5.496  &       91      & (7)& 90\\
Gl~880   &      4.130   &       0.58  & -4.744  &       37      & (7)& 28\\
Gl~382   &      4.170   &       0.53  & -4.655  &       22      & (7)& 25\\
Gl~514   &      3.990   &       0.52  & -4.879  &       30      & (7)& 35\\

\noalign{\smallskip}
\hline
\noalign{\smallskip}
\end{tabular*}
\end{table}

\section{$R^\prime_{HK}$ for the HARPS M dwarf sample}
\label{sec:Rhk_harps_sample}

We computed $R^\prime_{HK}$ for 403 M dwarfs observed by HARPS for 
planet searches \citep{2012A&A...546A..27B,2013A&A...549A.109B}. 
To estimate $C_{cf}$ and $R_{phot}$ , we used the V$-$K relation for 
Eqs.~\ref{eq:ccf_sol} and \ref{eq:rphot_sol} when a measurement of that 
color index was available in the literature, and we backed up to using 
the B$-$V relation for Eqs.~\ref{eq:ccf_sol} and \ref{eq:rphot_sol} when 
only the color was available.

Table~\ref{tab:caiihkharps} lists the median of 
the individual $R^\prime_{HK}$ measurements for each star, which
was used for the following analysis. The most active M dwarfs are 
fast rotators and, as such, were rejected from the planet search programs 
by the projected rotational velocity cut mentioned in Sect.~\ref{sec:obs}. 
Our sample is thus biased against the most active M dwarfs (but it nevertheless 
includes some of them) and representative of quiet and moderately active M dwarfs. 
Accordingly, our sample cannot be used to estimate the fraction of active 
stars among the M dwarfs. Tthis is not, however, a limitation to study the correlation 
between stellar parameters and stellar activity.

\subsection{$R^\prime_{HK}$ and metallicity}
\label{subsec:metallicity}

\begin{figure}[t]
\centering
\resizebox{\hsize}{!}{\includegraphics{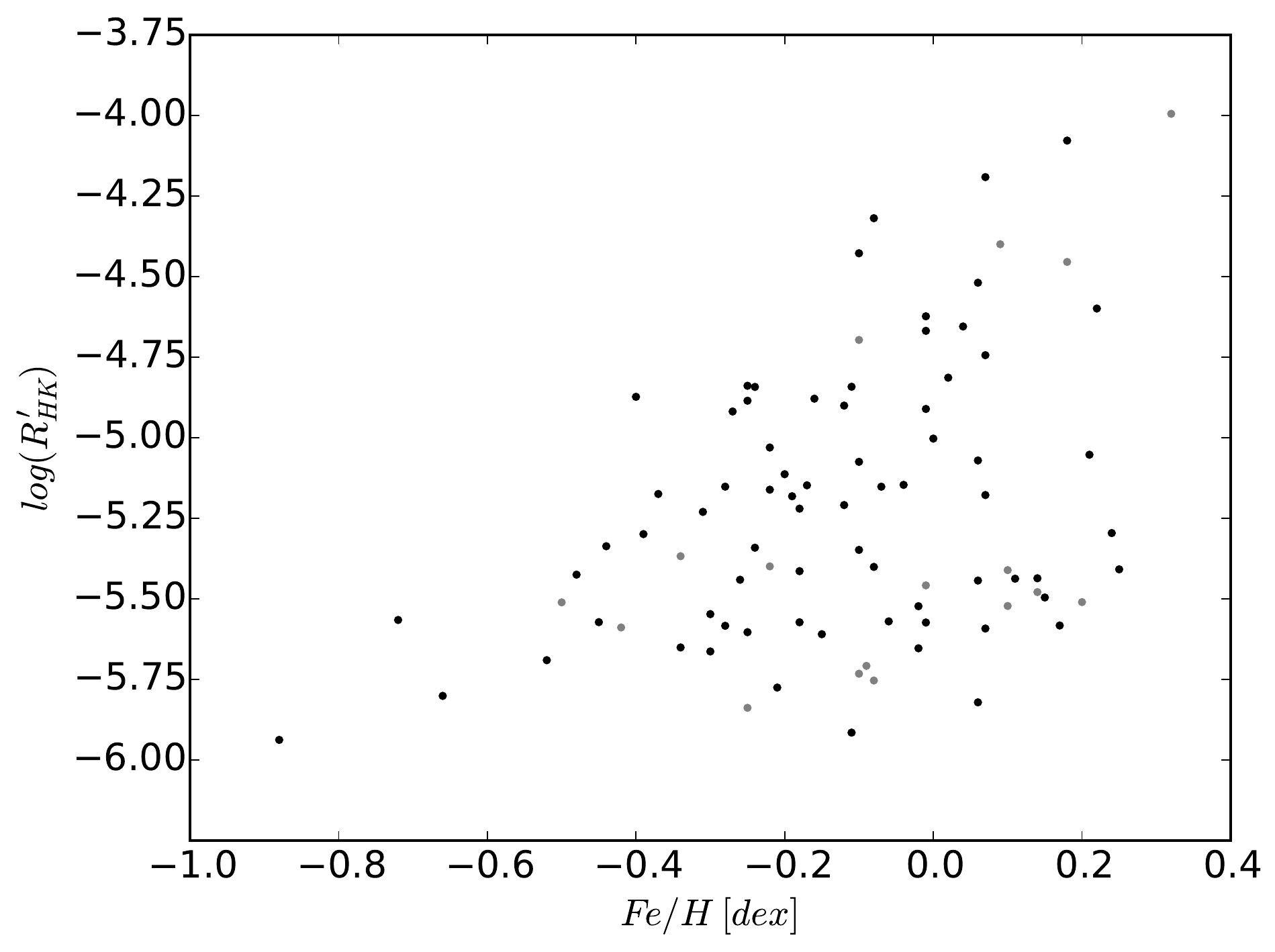}}
\caption{$log(R^\prime_{HK})$ as a function of metallicity. Black dots 
represent stars with at least six spectra that have S/N$_{VR,\,pix}\geq$5, 
while gray dots depict targets that do not match that condition. 
The lower envelope has no obvious trend with metallicity. }
\label{fig:Rhk_FeH}
\end{figure}

We first examined $R^\prime_{HK}$ against metallicity (Fig.~\ref{fig:Rhk_FeH}). 
Metallicity affects $R^\prime_{HK}$ measurements of the warmer solar-type stars, 
because, at a given $T_{eff}$, metal-poor stars have weaker fluxes in the V 
and R pseudo-continuum pass-bands, and therefore higher S values for the
same \ion{Ca}{ii} H and K flux; metallicity also affects the $C_{cf}$ 
factor, which is derived from the flux in the visual band, which in turn 
is sensitive to metallicity. 
\citet{2011arXiv1107.5325L} noticed from a 
linear trend in the lower envelope of a $R^\prime_{HK}$ versus $Fe/H$ 
diagram that $R^\prime_{HK}$  systematically decreases as $Fe/H$ increases,
and find that taking into account the variation of the bolometric flux 
as a function of $Fe/H$ eliminates that trend (their Fig. 3). 
Fig.~\ref{fig:Rhk_FeH} demonstrates that the lower envelope of the M dwarf
diagram shows no such trend, and we therefore include no metallicity 
term in our $R^\prime_{HK}$ calibration. One can also note that the active 
M dwarfs cluster in the metal-rich side of the diagram, as qualitatively 
expected from combination of the age-metallicity correlation and the 
decreasing chromospheric activity of older stars. 
Solar-metallicity M dwarfs of our sample nonetheless show 
$\log{R^\prime_{HK}}$ ranging from -6.0 to -4.25 while their FGK counterparts 
are less dispersed with $\log{R^\prime_{HK}}$ ranging from -5.1 to -4.65. 
Knowing that the most active stars have been rejected in our M dwarfs sample, 
the $\log{R^\prime_{HK}}$ dispersion for these stars might even be stronger. 
This could originate from the longer spin-down timescales of M dwarfs and, 
therefore, it takes a longer time for all the stars to converge to the sequence 
of the lower rotators.

\subsection{Dispersion of the $R^\prime_{HK}$ epoch measurements}
\label{subsec:dispersion}

The dispersion of the individual $R^\prime_{HK}$ measurements of a star contains 
contributions from true stellar variability, instrumental systematic error, and noise. 
As discussed in Sec.~\ref{sec:Rhk_accuracy}, the latter can 
often be dominant for M dwarfs, due to their low flux at the blue wavelengths 
of the \ion{Ca}{ii} H and K. To minimize this effect, we restrict discussion on 
variability to stars for which at least six $R^\prime_{HK}$ measurement 
have S/N$_{VR,\,pix}\geq$5 (see Sec.~\ref{sec:Rhk_accuracy} for more details).

\begin{figure}[t]
\centering
\resizebox{\hsize}{!}{\includegraphics{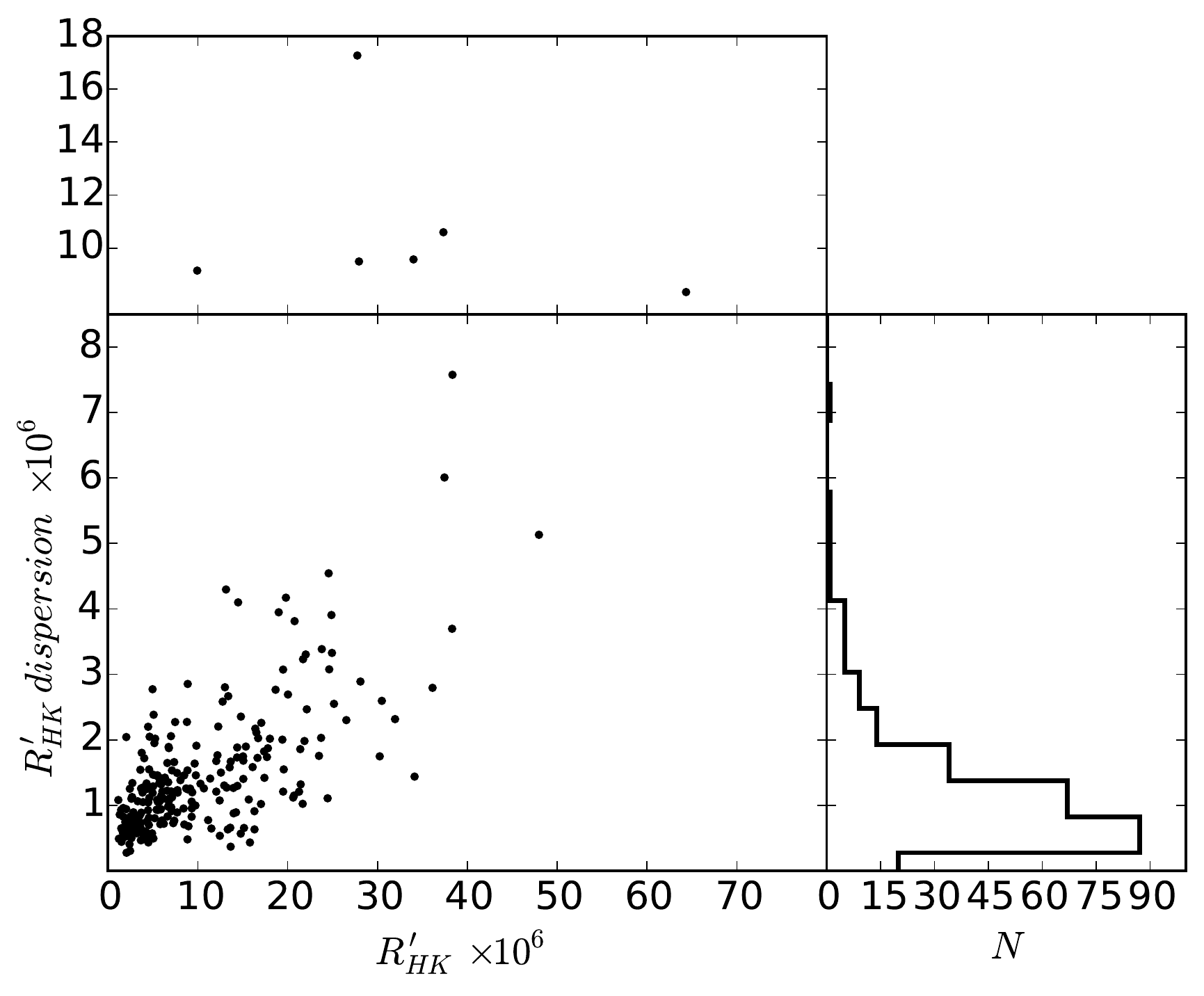}}
\caption{Average $R^\prime_{HK}$-index against the dispersion of its 
individual measurement for the 248 M dwarfs that satisfy the 
selection criteria described in the text. 
For clarity, the upper and lower panels separately display the
most variable stars and the bulk of the distribution with different
scales. The right-side panel shows a histogram of the $R^\prime_{HK}$ 
dispersion.}
\label{fig:Rhk_dispersion}
\end{figure}

Figure~\ref{fig:Rhk_dispersion} displays the median $R^\prime_{HK}$ against
its dispersion and shows that the more active stars are more variable,
as is also true for GK-dwarfs \citep[e.g.,][]{2011arXiv1107.5325L}. 
Many of the most active stars in Figure~\ref{fig:Rhk_dispersion} are 
known flare-stars, including Gl~551 (Proxima Centauri), Gl~54.1, 
Gl~729, GJ~3379, GJ~234AB, and GJ~3148A. The most likely $R^\prime_{HK}$ 
dispersion for an M dwarf is $1.2 \times 10^{-6}$, while the peak of the 
distribution is located at $0.6 \times 10^{-6}$, three~times higher than
the $0.2 \times 10^{-6}$ for G-K-dwarfs, and the M dwarf histogram
is also broader. This may stem from the slower spin-down of the M dwarfs, 
although residual instrumental effects could perhaps contribute.

Such strong intrinsic variability of the M dwarfs activity has 
often been pointed out, but few have been quantified until now. Such variability 
also concerns quiet M dwarfs.

\subsection{Activity as a function of stellar mass}
\label{subsec:mass}

\begin{figure}[t]
\centering
\resizebox{\hsize}{!}{\includegraphics{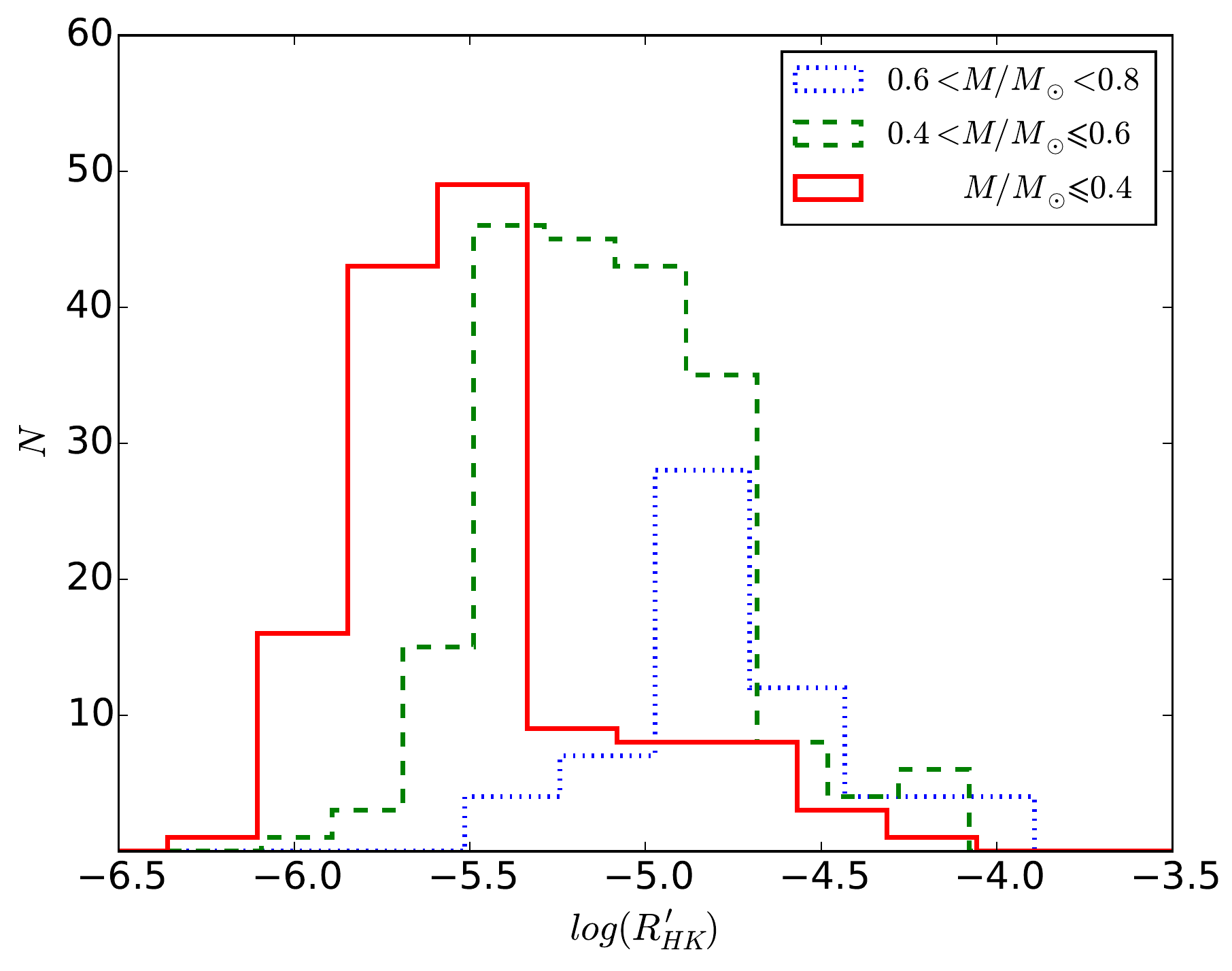}}
\caption{Histograms of the median $log(R^\prime_{HK})$ for three non-overlapping
M dwarf mass bins. The 403~M dwarfs used in the histograms divide into 138 
that are less massive than 0.4M$_\odot$, 206 with masses between 0.4M$_\odot$ 
and 0.6M$_\odot$, and 59 above 0.6M$_\odot$. Less massive M dwarfs are, on 
average, less active.}
\label{fig:Rhk_distr}
\end{figure}

We now turn our attention to the relationship between magnetic activity 
and stellar mass, restricting the discussion to stars with a well 
measured parallax ($\delta \pi / \pi < 0.1$) and for which
a mass can therefore be inferred from the \citet{2000A&A...364..217D} 
mass vs $M_K$ relation.
We first consider three mass bins, M/M$_\odot$$\leq$0.4, 
0.4$<$M/M$_\odot$$\leq$0.6, and 0.6$<$M/M$_\odot$$<$0.8. 
Fig.~\ref{fig:Rhk_distr} presents histograms of the median 
$log(R^\prime_{HK})$ for each bin, and shows that activity
level most likely decreases with stellar mass. The histogram for the highest
mass bin peaks at $log(R^\prime_{HK})$~=~-4.84, slightly higher than 
the -4.95 observed for G-K dwarfs \citet{2011arXiv1107.5325L}, that 
for the intermediate mass bin peaks at approximately -5.19, while that for the 
lowest-mass bin peaks at -5.47 but displays a strong tail of more active 
stars retained from our sample selection.

\begin{figure}[t]
\centering
\resizebox{\hsize}{!}{\includegraphics{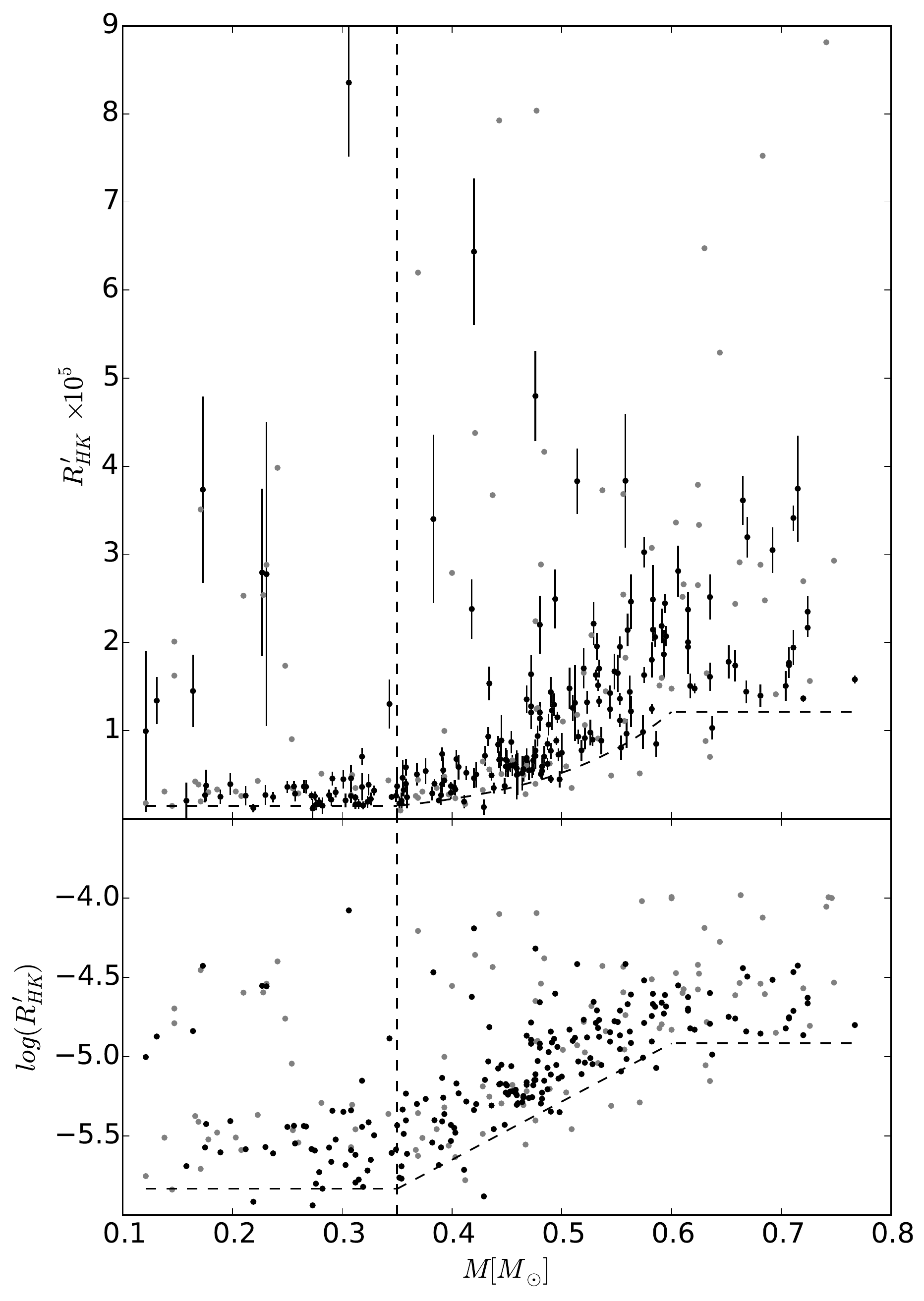}}
\caption{Median $R^\prime_{HK}$ as a function of stellar mass, with the
vertical errorbar representing the dispersion of the individual measurements
(upper panel). $log(R^\prime_{HK})$ as a function of stellar mass (lower panel).
Black dots represent targets with at least six measurements with 
S/N$_{VR,\,pix}>$5, while gray dots represent the rest of the stars.
Activity decreases with stellar mass down to approximately the 
mass for full convection (0.35M$_\odot$, marked by the vertical dashed 
line). 
Interestingly, M dwarfs departing further from the lower envelope 
(dashed curve) are those which present higher variability.}
\label{fig:Rhk_mass}
\end{figure}

For an unbinned view, Fig.~\ref{fig:Rhk_mass} displays $R^\prime_{HK}$ 
as a function of stellar mass. Both the lower envelope and the mode
of the  $R^\prime_{HK}$ distribution are approximately flat above 
$\sim 0.6M_\odot$, decrease with mass between $\sim$ 0.6$M_\odot$ and 
$\sim$0.35M$_\odot$, and flatten again below $\sim$0.35M$_\odot$. 
The later break approximately coincides with the transition from partially 
to fully convective stars \citep{2000ARA&A..38..337C}, and could potentially 
reflect a change at this transition where the $\alpha\Omega$ 
dynamo vanishes and a change is observed in the topology of the large-scale 
component of their magnetic field \citep{2008MNRAS.390..545D,2008MNRAS.390..567M}.

For the earlier spectral types, our findings are consistent with those of 
\citet{2013A&A...549A.117M}. They found that \ion{Ca}{ii} activity 
rises over 1.1$\leq$B$-$V$\leq$1.3 (K5 to K7-dwarfs) and decreases over 
1.3$\leq$B$-$V$\leq$1.5 (M0 to M3-dwarfs), but could not establish 
whether or not that decrease continues to later spectral types since their 
sample does not extend beyond M3. Fig.~5 in \citet{2010AJ....139..504B} 
shows a similar behavior of $L_{Ca}/L_{Bol}$ against spectral type for 
M dwarfs, albeit less clearly.

$H\alpha$ shows a similar behavior although the decline starts at mid-M dwarfs. 
$L_{H\alpha}/L_{bol}$ remains approximately constant for spectral types M0 to M5 and only 
starts to decline at M5-M6 \citep{2004AJ....128..426W}. 
This is not inconsistent, since \ion{Ca}{ii} H and K and H$\alpha$ trace different 
chromospheric heights.
Their emissions do positively correlate for very active stars, 
but not for intermediate or weak activity stars 
\citep{2006PASP..118..617R,2009AJ....137.3297W}.

\section{Conclusions and summary}
\label{sec:conclusions}

We use high-resolution spectra of M dwarfs observed in HARPS planet-search
programs to analyze their \ion{Ca}{ii} H and K magnetic activity and
examine how it varies with stellar rotation period, stellar mass, and
color. For this purpose, we extend the  B$-$V photometric calibrations of
the bolometric $C_{cf}$ and photospheric $R_{phot}$ factors used in the
computation of the $R^\prime_{HK}$-index to B$-$V=1.90 (spectral type M6).
We also derive alternative, and preferred, I$-$K and V$-$K calibrations
of these two factors.

We calibrated the $C_{cf}$ relationship in a purely 
empirical way, without the use of synthetic spectra, through the integrated 
flux in the V and R control bands, and the bolometric flux determined by the 
integrated flux in the visual band and the bolometric correction 
\citep{1996ApJ...469..355F,2001ApJ...548..908L}. On the contrary, the $R_{phot}$ 
relationship is calibrated in using a synthetic spectrum that replaces a narrow 
window (2 $\AA$) around \ion{Ca}{ii} H and K lines of observed spectra (chromosphere+photosphere).
While an extrapolation of the \citet{1984A&A...130..353R} $C_{cf}$ vs B$-$V 
relation agrees reasonably well with our new relation for B$-$V$<$1.6, 
we find that extrapolating the \citet{1982A&A...107...31M} or 
\citet{1984ApJ...279..763N} , as done by
\citet[e.g.][]{2002MNRAS.332..759T,2006MNRAS.372..163J}
overestimates $R^\prime_{HK}$ values by factors of two to three for mid- to
late-M dwarfs.

The $log(R^\prime_{HK})$ vs. $log(P_{rot})$ diagram of M0-M6 dwarfs displays
two distinct regimes, with  $log(R^\prime_{HK})$ saturated for $P < 10\ d$
(or a Rossby number of $\approx0.1$ for an assumed $\tau_c=70\ d$
convective turnover timescale) and then decreasing as a linear function
of  $log(P_{rot})$ for longer rotational periods.
These two regimes similarly appear in other tracers of magnetic activity
\citep[such as $L_X/L_{bol}$ or $Bf$, ][]{2007AcA....57..149K,2009ApJ...692..538R}.
$R^\prime_{HK}$, $L_X/L_{bol}$, and $Bf$ all depend similarly on stellar
rotation in the unsaturated regime, as expected if rotation drives
all magnetic activity.
The stellar rotation can be estimated from $L_X/L_{bol}$
\citep[e.g.,][]{2007AcA....57..149K}, although $R^\prime_{HK}$ has the
 potential to derive $P_{Rot}$ for quieter stars, where X emission
is eventually not detected.

The lower envelope of the 
\ion{Ca}{ii} H and K activity varies with mass over the M dwarf range.
The basal level of the $R^\prime_{HK}$ distribution decreases
with the lower envelope and mode of the $R^\prime_{HK}$ distribution decreases
with stellar mass down to $M\sim0.35M_\odot$ and flattens below that mass,
which coincides with the transition from partially to fully convective stars.
It will be difficult to confirm if another basal coronae and chromosphere emission decrease with mass for M dwarfs, $L_{H_{\alpha}}/L_{BOL}$ or $L_X/L_{BOL}$ not being determined for the quietest stars.

Besides insight into surface magnetic fields, Eq.~(\ref{eq:R_Prot_relation})
provides information on $P_{rot}$ (with a typical accuracy of 8 days)
from a measurement of $R^\prime_{HK}$ that
can be obtained from a single high-resolution spectrum. This has
significant practical importance in the context of extra-solar planet
searches, where stellar activity modulated by rotational visibility
is an important source of false-positives
\citep[e.g.,][]{2007A&A...474..293B, 2014Sci...345..440R}.
A good estimate of $P_{rot}$ from Eq.(~\ref{eq:R_Prot_relation}) can
thus retire false positive worries when the potential signal is
sufficiently removed from the estimated stellar-rotation period  
and its harmonics \citep{2011A&A...528A...4B}, and will intensify
such worries when it is not.

\begin{acknowledgements}
Based on observations made with the HARPS instrument on 
    the ESO 3.6 m telescope under program IDs 072.C-0488(E), 
    183.C-0437(A), 072.C-0488, 183.C-0972 and 083.C-1001 at 
    Cerro La Silla (Chile).
The authors acknowledge Nad\`ege Meunier and Eduardo Mart\'in for theirs precious comments.
N. A.-D. acknowledges support from CONICYT Becas-Chile 72120460.
X. B., X. D., and T. F. acknowledge the support of the French Agence
Nationale de la Recherche (ANR), under the program ANR-12-BS05-0012
Exo-atmos and of PNP (Programme national de plan\'etologie).
This work has been partially supported by the Labex OSUG@2020.
X.B. acknowledges funding from the European Research Council under
the ERC Grant Agreement n. 337591-ExTrA.
\end{acknowledgements}

\bibliographystyle{aa}
\bibliography{Rhk_v5}

\section{Tables}

\begin{table}[p]
\centering
\caption{Median values for $C_{cf}$ and for $R_{phot}$ for M dwarfs. N is the number 
of spectra satisfying restrictions described in Sect.~\ref{sec:bol_cor}. $C_{cf}$  
values are derived from Eq.~(\ref{eq:ccf3}); $R_{phot}$ is obtained with 
equations~(\ref{eq:s_harps}), (\ref{eq:s_harps_wright_lin}), (\ref{eq:Rhk}), and 
(\ref{eq:ccf_sol}) V$-$K relation after the reversal Ca \textrm{\small II} H and K 
emission was replaced by the corresponding BT-Settl spectrum.}
\label{tab:targets_ccf_color_M}

\end{table}

\begin{table*}[t]
\centering
\caption{$C_{cf}$  and median $R_{phot}$ for standards. 
$C_{cf}$ values are derived from \citet{1982A&A...107...31M}'s relation. 
$R_{phot}$ is obtained in the same way as is described in table~\ref{tab:targets_ccf_color_M}.}
\label{tab:targets_ccf_color_std}
}

\end{document}